\documentclass[12pt]{article} 
\usepackage{epsfig,ldd_art,latexsym}
\input{epsf.tex}
\newcommand{\One}{1\kern-4.5pt1}

\begin{document}

\addtolength{\baselineskip}{0.20\baselineskip}

\rightline{ SWAT/00/260}
\rightline{ DESY 00-083}

\hfill June 2000
\vspace{24pt}

\centerline{\bf NUMERICAL STUDY OF DENSE ADJOINT MATTER}
\centerline{\bf IN TWO COLOR QCD} 


\vspace{18pt}

\centerline{\bf Simon Hands$^a$, Istv\'an Montvay$^{b,c}$, Susan
Morrison$^a$
Manfred Oevers$^d$,} 
\centerline{\bf Luigi Scorzato$^a$ and Jonivar Skullerud$^b$}

\vspace{15pt}

\centerline{\sl $^a$Department of Physics, University of Wales Swansea,}
\centerline{\sl Singleton Park, Swansea SA2 8PP, U.K.}

\centerline{\sl $^b$Theory Division, DESY, Notkestrasse 85, D-22603 Hamburg, 
Germany.}

\centerline{\sl $^c$Center for Computational Physics, University of Tsukuba,}
\centerline{\sl
1-1-1 Tennodai, Tsukuba-shi, Ibariki-ken 305-8577, Japan.}

\centerline{\sl $^d$Department of Physics and Astronomy, University of Glasgow,}
\centerline{\sl Glasgow G12 8QQ, U.K.}

\vspace{12pt}


\centerline{{\bf Abstract}}

\noindent
{\narrower 
We identify the global symmetries of SU(2) lattice gauge theory
with $N$ flavors of staggered fermion in the presence of a quark
chemical potential $\mu$, for fermions in both fundamental
and adjoint representations, and anticipate likely patterns of symmetry
breaking at both low and high densities. Results from numerical simulations
of the model with $N=1$ adjoint flavor on a $4^3\times8$ lattice 
are presented, using both hybrid Monte Carlo 
and Two-Step Multi-Boson algorithms. It is shown that the sign of the fermion 
determinant starts to fluctuate once the model enters a phase
with non-zero baryon charge density. HMC simulations are not ergodic in 
this regime, but TSMB simulations
retain ergodicity even in the dense phase, and in addition appear to 
show superior decorrelation.
The HMC results for the equation of state and the pion mass
show good quantitative
agreement with the predictions of chiral perturbation theory, which should hold
only for $N\ge2$. The TSMB results incorporating
the sign of the determinant support a delayed onset transition,
consistent with the pattern of symmetry breaking expected for $N=1$.}



\vfill

\newpage

\section{Introduction}

In QCD it is believed that as
the baryon density rises the degrees of freedom providing the
most suitable description change
from being ``hadronic'', ie.\ composite states such as protons and neutrons, to
being ``partonic'', ie.\ quarks and gluons. This change may well be signalled
by a phase transition as the appropriate thermodynamic variable, the baryon
chemical potential $\mu$, is raised.
Recent theoretical speculation \cite{BL,ARW,ARW2} 
suggests that the ground state of
strongly-interacting quark matter at high density
may be more exotic than initially
thought, for instance existing in a superconducting and/or
superfluid state due to the
condensation of diquark pairs at the Fermi surface via a BCS instability.
In such a state the SU(3) color gauge group is spontaneously broken by a
dynamical Higgs mechanism; in the language of condensed-matter physics this is
the Meissner effect. As well as being of intrinsic theoretical
interest, the behaviour of strongly-interacting matter at extreme densities
is of fundamental importance both to nuclear physics and
in understanding compact astrophysical objects
such as neutron stars. 

Unfortunately, it is difficult to apply
the most reliable calculational tool for QCD,
lattice gauge theory, directly to this
problem. The reason is that once $\mu\not=0$, the anti-hermitian property
of the Euclidian Dirac operator $D$ governing the motion of the
quarks $q$ and anti-quarks $\bar q$
in the presence of the color field is spoiled, with the result that the
functional measure
$\mbox{det}M$, with $M\equiv D(\mu)+m$ where $m$ is the quark mass, 
is no longer positive definite.
To apply the normal 
Monte Carlo method of importance sampling the path integral,
the determinant must be split into a modulus and a phase; importance sampling
is then done with respect to a measure $\vert\mbox{det}M\vert$, and 
$\mbox{arg(det}M)$ is incorporated with the observable:
\begin{equation}
\langle O\rangle = {{\langle\langle O\,\mbox{arg(det} M)\rangle\rangle}\over
                    {\langle\langle \mbox{arg(det} M)\rangle\rangle}},
\label{eq:obs}
\end{equation}
where $\langle\langle\ldots\rangle\rangle$ denotes the expectation value with 
respect to the positive real measure. Now, the denominator
of (\ref{eq:obs}) is in effect a ratio of the partition functions of two 
different theories, one the true theory and the other an artificial
one with a positive real measure: it should thus scale as $\exp(-\Delta F)$, 
where the free energy difference between the theories $\Delta F$  is
an extensive quantity.
The number of states to be
sampled before estimators of observables converge therefore
in general rises exponentially with the
system volume.  This is the origin of the notorious `sign problem'. 

We can gain physical insight into the sign problem by considering 
standard QCD simulation algorithms such as the hybrid Monte
Carlo (HMC)
algorithm, which  use a positive definite measure $\mbox{det}M^\dagger M$.
The $M^\dagger$ has the effect of introducing ``conjugate quarks'' $q^c$
transforming
in the conjugate representation of the color group
\cite{Gocksch}. In general the presence
of light gauge invariant bound
$qq^c$ states carrying net baryon number in the spectrum
results in unphysical behaviour for $\mu\not=0$,
eg.\ a premature ``onset'' transition
between the vacuum and nuclear matter at $\mu_o\sim O(m_\pi)$, the pion mass,
which if chiral symmetry is spontaneously broken is much smaller than 
the constituent quark mass scale at which the transition is expected.
There are, however, two strongly-interacting model theories where conjugate
quarks can be tolerated. First consider the Nambu--Jona-Lasinio
(NJL) model, long used as an effective
theory for strong interactions; for sufficiently strong coupling
it displays chiral symmetry breaking, signalled
by a non-vanishing condensate $\langle\bar q q\rangle$ and the development of a
constituent quark mass 
greatly exceeding the current quark mass $m$, together
with a triplet of light mesonic  (ie.\ $q\bar q$)
pion states via Goldstone's theorem. Conjugate quarks in 
numerical simulations of the NJL model
are harmless because the diagrams responsible for the tight binding
and small mass of the pions are only accessible in $q\bar q$ channels
\cite{BHKLM}.

Next, consider non-abelian gauge theory in which 
the color group is SU(2) rather
than the
physical SU(3). Over the years ``Two Color QCD'' 
has been used to study the strong interaction
in a variety of different contexts,
the main motivation being that the computer effort required is
appreciably less. Once quarks are introduced, however, important physical
differences become apparent. Because the matter representations are all either
real or pseudoreal, there is no gauge quantum number distinguishing $q$ from
$\bar q$, resulting in enhanced global flavor symmetries in which both
$q\bar q$ meson and $qq$ baryon states appear in the same multiplet
\cite{Peskin}, and
chiral $\langle\bar qq\rangle$ and diquark $\langle qq\rangle$ condensates
are related by global rotations \cite{early,HKLM,KSTVZ}. The
lightest baryon is therefore degenerate with the pion, and the onset chemical
potential vanishes as $\mu_o\propto\surd m$ in the chiral limit.
The same features ensure $\mbox{det}M(\mu)$ is real, and hence the theory
simulable using standard algorithms \cite{HKLM,Lombardo}.

Our motivations for studying Two Color QCD with adjoint quarks are twofold.
Firstly, as we shall demonstrate, the pattern of symmetry breaking anticipated
for gauge theories with quarks in real or pseudoreal representations of the
gauge group differs between 
continuum \cite{KSTVZ} and staggered lattice fermions
\cite{HKLM}. There are good reasons, therefore, for considering 
the model with adjoint rather than fundamental
quarks to be the most `QCD-like' of the lattice models;
in particular it has gauge invariant spin-$1\over2$ states in its spectrum, 
for $N=1$ quark flavors no baryonic Goldstone modes are expected, and it
could potentially have a superconducting ground state at high density. 
Secondly, for an odd number of flavors the functional measure, though real,
is not positive definite. The model thus has a potential sign problem of 
a simpler form than QCD in that only sectors having opposite sign,
rather than a continuum of phases, need be considered. It may thus be feasible 
to make progress using standard means, or at least expose physical distinctions
between the two sectors.

Apart from the sign problem, which may almost be considered a problem of
principle in the study of non-zero chemical potential, there are practical
problems in using the HMC algorithm once $\mu\not=0$. For instance, 
the numerical effort required to invert the matrix $M(\mu)$ rises considerably
as $\mu$ increases, due to the proliferation of complex eigenvalues with small
modulus. This has encouraged us to study the performance of an alternative
approach, the Two-Step Multi-Bosonic (TSMB) algorithm, 
in which $M$ is not inverted
at all,
but the effect of $\mbox{det}M$ incorporated
by local Monte Carlo simulation over many
auxiliary boson fields \cite{LUSCHER}. Our conclusion is that the TSMB
algorithm may be by far the more effective approach, in terms of the cost to
produce decorrelated configurations, in the high density phase. Moreover for
the class of problem in which $\mbox{det}M$ is real but not positive our
results suggest the HMC algorithm in its simplest form is not ergodic in the
same region of parameter space, since it fails to change the sign of the
determinant. The TSMB algorithm does not share this problem.

The remainder of the paper is organised as follows.
In Section 2 we outline the
global symmetries of the lattice model, and review the expected breaking 
pattern, for both adjoint and fundamental quarks. Operators 
for possible diquark condensates 
which form at high baryon density are discussed. We present
a proof that $\mbox{det}M$ is real and positive for fundamental lattice quarks,
but only real for adjoint. Finally issues concerning the continuum limit,
which is problematic, are discussed. In section 3 we discuss both HMC and 
TSMB algorithms, and analyse their performance by considering the eigenvalue
spectrum of $M$. It will be made clear that both sign problem and ergodicity 
are issues of practical importance once $\mu>\mu_o$, 
and that the TSMB approach is better
suited to tackling them.
In section 4 we present results of simulations using both algorithms. HMC
simulations, confined to the sector of positive determinant,
show clear evidence of an onset phase transition at $\mu_o\simeq m_\pi/2$
to a ground state with a 
non-zero density of baryon charge. The results moreover are in excellent
agreement with analytic predictions obtained using chiral perturbation theory
\cite{KSTVZ}, which might be expected to hold for more than one quark flavor. 
By way of contrast the TSMB simulations, 
which take the determinant sign
into account, show evidence that this onset transition is thereby delayed.
Our conclusions and future plans are briefly outlined in Section 5.

\section{Two Color QCD on the Lattice} 

\subsection{Formulation and Symmetries at Low Density}
\label{subs:form}

In this section we review the formulation
and symmetries of Two Color QCD with staggered
lattice fermions in the presence of a chemical potential $\mu$, 
and compare them to those of the corresponding continuum models.
In this way we hope to motivate the study of the model with $N=1$ flavor of
adjoint quark as the most `QCD-like' of the possibilities. The fermionic 
part of the lattice action is as follows: 
\begin{equation}
S=\sum_{x,y}\bar\chi^p(x)D_{x,y}[U,\mu]\chi^p(y)+
m\bar\chi^p(x)\delta_{x,y}\chi^p(y)
\equiv\sum_{x,y}\bar\chi^p(x)M_{x,y}[U,\mu]\chi^p(y)\label{eq:s0},
\end{equation}
where the index $p$ runs over $N$ flavors of staggered quark, 
and $D$ is given by
\begin{eqnarray}
D_{x,y} = &{1\over2}&\sum_{\nu\not=0}\eta_\nu(x)
\left(U_\nu(x)\delta_{x,y-\hat\nu}
        -U_\nu^\dagger(y)\delta_{x,y+\hat\nu}\right)\cr+
&{1\over2}&\eta_0(x)\left(e^\mu U_0(x)\delta_{x,y-\hat0}
        -e^{-\mu}U_0^\dagger(y)\delta_{x,y+\hat0}\right)\label{eq:s1}.
\end{eqnarray}
The $\chi, \bar\chi$ are single spin component Grassmann objects, 
and the phases $\eta_\mu(x)$ are defined to be $(-1)^{x_0+\cdots+x_{\mu-1}}$.

In the case of fundamental quarks, the link matrices $U_\mu$ are complex
$2\times2$ matrices acting on isodoublet $\chi,\bar\chi$, and may be 
parametrised in terms of 3 real numbers $\alpha_i$ 
as $U=\exp(i\alpha_i\tau_i)$, 
where $\tau_i$ are the Pauli matrices. Note that $\tau_2U\tau_2=U^*$.
For adjoint quarks, the same group elements may be 
represented by real $3\times3$ orthogonal matrices $O$ acting on
isotriplet $\chi, \bar\chi$, given by
\begin{equation}
O_{ij}={1\over2}\mbox{tr}(\tau_iU\tau_jU^\dagger). 
\end{equation}
In terms of the $\alpha_i$, $O=\exp(2i\alpha_it^i)$, where in this
representation the generators $(t^i)_{jk}=-i\varepsilon_{ijk}$ are hermitian,
pure imaginary, and antisymmetric. For notational convenience we will
continue to write the link variables as $U_\mu$ in either case.

On integration over $\chi$ and $\bar\chi$ the effective action 
$\exp(-S_{eff})=\mbox{det}^NM$ is obtained. 
Unlike three-color QCD with $\mu\not=0$, 
$\exp(-S_{eff})$ is real, since in the fundamental case $\mbox{det}M
=\mbox{det}\tau_2M\tau_2=\mbox{det}M^*$, while in the adjoint case 
$M$ is manifestly real.

In the chiral limit $m=0$, the action has two manifest
global symmetries:
\begin{eqnarray}
\mbox{U}(N)_e:\;\;\; \chi_e & \mapsto & P\chi_e\;
;\bar\chi_o\mapsto\bar\chi_o P^\dagger
\;\;\;P\in \mbox{U}(N) \nonumber\\
\mbox{U}(N)_o:\;\;\; 
\chi_o &\mapsto& Q\chi_o\;;
\bar\chi_e\mapsto\bar\chi_e Q^\dagger\;\;\;
Q\in \mbox{U}(N),
\end{eqnarray}
the $e/o$ subscripts denoting fields on even and odd
sublattices respectively.
However, it is straightforward to rearrange (\ref{eq:s0},\ref{eq:s1})
using the Grassmann nature of $\chi,\bar\chi$ and the fact that
$\eta_\mu(x\pm\hat\mu)=\eta_\mu(x)$ to rewrite the action in this limit as
\begin{eqnarray}
S={1\over2}\sum_{x even, \nu}\eta_\nu(x)
\biggl[\bar X_e(x)&\left(\matrix{e^{\mu\delta_{\nu,0}}&\cr
              &e^{-\mu\delta_{\nu,0}}\cr}\right)&
                           U_\nu(x)X_o(x+\hat\nu)-  \cr
\bar X_e(x)&\left(\matrix{e^{-\mu\delta_{\nu,0}}&\cr
              &e^{\mu\delta_{\nu,0}}\cr}\right)&
U_\nu^\dagger(x-\hat\nu)X_o(x-\hat\nu)\biggr]
\end{eqnarray}
where the fields $X,\bar X$ are given by
\begin{equation}
\bar X_e=(\bar\chi_e,-\chi_e^{tr}\tau_2)\;\;:\;\;X_o=\left(\matrix{
\chi_o\cr-\tau_2\bar\chi_o^{tr}\cr}\right)
\end{equation}
for fundamental quarks and
\begin{equation}
\bar X_e=(\bar\chi_e,\chi_e^{tr})\;\;:\;\;X_o=\left(\matrix{
\chi_o\cr\bar\chi_o^{tr}\cr}\right)
\end{equation}
in the adjoint case.
In the limit $\mu\to0$ the 
U($N$)$_e\otimes$U($N$)$_o$ symmetry thus enlarges to U($2N$):
\begin{equation}
X_o\mapsto VX_o\;\;\bar X_e\mapsto\bar X_eV^\dagger\;\;\;V\in {\rm U}(2N).
\label{eq:U2N}
\end{equation}
Note that the U($2N$) group emerges
because exact symmetries are non-anomalous in lattice formulations; 
for a continuum model
with $N_f$ flavors the analogous symmetry enlargement is
SU($N_f)_L\otimes$SU($N_f)_R\otimes$U(1)$_B\to$SU($2N_f$) \cite{Peskin}.

For a non-abelian gauge theory we expect spontaneous
chiral symmetry breaking
to occur at low density, signalled by the appearance of a chiral condensate
$\langle\bar\chi\chi\rangle\not=0$, which in the first instance we will
consider to have the same form as the bare quark mass term in (\ref{eq:s0}).
To determine the pattern of symmetry breaking it is helpful to recast the
condensate in terms of $X,\bar X$: we find
\begin{equation}
\bar\chi\chi={1\over2}\left[
\bar X_e\left(\matrix{&\One\cr\One&\cr}\right)\tau_2\bar X_e^{tr}+
X_o^{tr}\left(\matrix{&\One\cr\One&\cr}\right)\tau_2X_o\right]
\label{eq:condf}
\end{equation}
in the fundamental case, and
\begin{equation}
\bar\chi\chi={1\over2}\left[
\bar X_e\left(\matrix{&\One\cr-\One&\cr}\right)\bar X_e^{tr}-
X_o^{tr}\left(\matrix{&\One\cr-\One&\cr}\right)X_o\right]
\label{eq:conda}
\end{equation}
for the adjoint. Here $\One$ denotes the $N\times N$ unit matrix. The residual
symmetry left unbroken by the condensate is that which leaves invariant 
respectively the symmetric/antisymmetric $2N\times2N$ form. 
For $\mu=0$ we thus find
\begin{equation}
\mbox{fundamental}:\;\;\;\mbox{U}(2N)\to\mbox{O}(2N)\;\;\;\;\;\;
\mbox{adjoint}:\;\;\;\mbox{U}(2N)\to\mbox{Sp}(2N).
\label{eq:latt_br}
\end{equation}
This is remarkable in that it is the opposite of the
breakdown in the continuum \cite{Peskin}:
\begin{equation}
\mbox{fundamental}:\;\;\;\mbox{SU}(2N_f)\to\mbox{Sp}(2N_f)\;\;\;\;\;\;
\mbox{adjoint}:\;\;\;\mbox{SU}(2N_f)\to\mbox{O}(2N_f).
\end{equation}
In effect the r\^oles of fundamental and adjoint representations are reversed
for staggered lattice fermions, a fact which has been noted several times
over the years \cite{U2N}. 

Next we identify the massless modes arising in the $m,\mu\to0$ limit as 
a result of Goldstone's theorem. For the fundamental case, the number of broken
generators predicted by (\ref{eq:latt_br})
is $N(2N+1)$. The case of $N=1$ fundamental quark has been analysed
in \cite{HKLM} by 
considering infinitesimal rotations of the condensate (\ref{eq:condf})
by $V_\delta=\One+i\delta\lambda$, with $\lambda$ one of the U(2) generators
$\{\One,\tau_i\}$, and identifying the Goldstone mode with the coefficient 
of $O(\delta)$. The three Goldstones thus found are the familiar mesonic 
$0^-$ pion
$\bar\chi\varepsilon\chi$ (where $\varepsilon(x)=(-1)^{x_0+x_1+x_2+x_3}$), 
and two scalar $0^+$ diquark states
$\chi^{tr}\tau_2\chi,\,\bar\chi\tau_2\bar\chi^{tr}$.
The $\pm$ superscripts here denote the symmetry of the state under the
following lattice `parity' symmetry:
\begin{eqnarray}
x=(x_0,x_1,x_2,x_3) &\mapsto& x^\prime=(x_0,1-x_1,1-x_2,1-x_3)\cr
\chi(x)\mapsto(-1)^{x_1^\prime+x_3^\prime}\chi(x^\prime)&;&
\bar\chi(x)\mapsto(-1)^{x_1^\prime+x_3^\prime}\bar\chi(x^\prime).
\label{eq:lattpar}
\end{eqnarray}
For $m\not=0$, the three states remain degenerate \cite{HKLM}, gaining
masses $m_\pi\propto\surd m$ in accordance with standard PCAC arguments.
In the case of adjoint quarks, the pattern (\ref{eq:latt_br}) predicts
$N(2N-1)$ Goldstones in general, and for $N=1$ the only Goldstone
mode is the $0^-$ mesonic pion. The corresponding analysis for the
continuum models has been performed for $N_f\ge2$
with some care in \cite{KSTVZ}; the Goldstone counts are found to be 
$N_f(2N_f-1)-1$ (fundamental) and $N_f(2N_f+1)-1$ (adjoint). 
Modulo the mode destroyed by the U(1) axial anomaly, the reversal of
fundamental and adjoint cases is clear. Another important distinction is that in
the continuum the Goldstone spectrum {\sl always} contains diquark
states; in this respect the $N=1$ lattice adjoint model is special.

Now consider the effect of increasing $\mu$
from zero. The chemical potential has the effect of promoting a ground state
containing baryonic matter, signalled by a non-zero
value for the baryon number density 
\begin{equation}
n={1\over2}\left\langle\bar\chi(x)\eta_0(x)[e^\mu U_0(x)\chi(x+\hat0)+
e^{-\mu}U_0^\dagger(x-\hat0)\chi(x-\hat0)]\right\rangle. 
\label{eq:n}
\end{equation}
At zero temperature
$n$ thus serves as an order parameter for an {\sl onset\/} phase transition
occuring at some $\mu_o$ separating the vacuum from a state containing matter.
A naive energetic argument would 
suggest that the onset transition should occur for a value of $\mu_o$ 
equal to 
the mass per baryon charge  of the lightest particle carrying 
non-zero baryon number. For the models discussed in the previous paragraph
in which some of the Goldstone modes are diquark states, those states will
be the lightest baryons in the spectrum. Generically then, we expect
$\mu_o\simeq m_\pi/2$ for most variants of Two Color QCD, which should be
contrasted with the much larger value 
$m_{nucleon}/3$ expected in physical QCD.
The exception is
the lattice adjoint model with $N=1$. The fact that the lightest baryons
are Goldstone modes and thus amenable to analysis by using Chiral
Perturbation Theory ($\chi$PT) on an effective sigma
model has been exploited in \cite{KSTVZ,KST} to calculate
both Goldstone spectrum and equation of state of the continuum 
models as functions of $\mu$; 
the principal result is the prediction $\mu_o=m_\pi/2$.

\subsection{Diquark Condensation at High Density}
\label{subs:diquark}

For a sufficiently high density of baryon charge, regardless of the nature
of the bound states in the low energy spectrum,
the fermionic nature of the quarks should ensure that the dominant degrees of
freedom are governed by a Fermi-Dirac distribution. At zero temperature, 
all states will be occupied up to the Fermi energy $E_F$, which coincides
with $\mu$ in the limit where inter-quark interactions can be neglected (for
an asymptotically free theory we expect this approximation to improve as
$\mu$ rises). The question now arises as to whether this simple description is
unstable with respect to condensation of diquark pairs situated at antipodal
points on the Fermi surface, resulting in an energy gap between the ground
state and the lowest spin-$1\over2$ excitation. For large $\mu$ this instability
is generic provided the quark-quark interaction is attractive; that this is
so for non-abelian gauge theories has been argued as arising from either gluon
exchange \cite{BL}, or instanton effects \cite{ARW}. 
Physically, diquark condensation implies that fermion number and/or
baryon charge is no longer
conserved, and the ground state is a superfluid. For physical QCD, diquark
pairs cannot be color singlet, so the condensation results in the phenomenon
of color superconductivity, rendering some or all of the gluons massive via a
dynamical Higgs mechanism \cite{ARW,ARW2}.

We now consider possible diquark condensates that might form in 
Two Color QCD. There are many possible diquark states that can be written down,
and in the absence of a detailed dynamical calculation we have to proceed by
making some {\it ad hoc\/} assumptions. We might imagine that the
wavefunction of the $qq$ condensate should ideally be gauge invariant, spacetime
scalar, and in the case of the lattice model, as local as possible in the 
$\chi$ fields, since non-local wavefunctions require the insertion of link
variables to maintain gauge invariance, whose fluctuations will weaken the
condensation. The most important consideration (and the only one which 
is inviolate \cite{ARW,HM}) is that the condensate respects the Pauli Exclusion
Principle, implying that the operator be antisymmetric with respect to
exchange of quantum numbers between the quarks.

For Two Color Lattice QCD with fundamental quarks, a diquark operator
which satisfies all of the above requirements is
\begin{equation}
qq_{\bf2}={1\over2}\left[\chi^{tr}(x)\tau_2\chi(x)+
\bar\chi(x)\tau_2\bar\chi^{tr}(x)\right].
\label{eq:qq2}
\end{equation}
In fact \cite{HKLM}, $qq_{\bf2}$ is related to the chiral condensate
$\bar\chi\chi$ by the U($2N$) symmetry (\ref{eq:U2N}), 
and as $\mu$ increases the chiral
condensate (\ref{eq:condf})
is in effect {\sl rotated\/} into the diquark one. There is, 
however, a physical distinction between high and low density phases; 
with $N$ set to 1 and $m,\mu>0$ the U(2) symmetry is reduced to U(1)$_B$ which 
is associated with conservation of baryon number. Condensation of $qq_{\bf2}$
spontaneously breaks this residual symmetry resulting in an exactly massless
Goldstone mode. Because of the change in the number of massless modes, there is
a true phase separation between either the vacuum or a low density normal phase,
and a high density 
superfluid phase. Preliminary Monte Carlo simulations have revealed
a transition as $\mu$ is increased \cite{HKLM}, and the associated condensation 
of $qq_{\bf2}$ has been confirmed \cite{MH}. A superfluid condensate at high
density is also
found in both fundamental and adjoint continuum models \cite{KSTVZ}; here 
the sigma
model approach permits a calculation of the spectrum and equation of state 
as a function of $\mu/m$, and predicts that $\langle qq\rangle$ and $n$
become non-zero at the same critical $\mu_o=m_\pi/2$, ie.\ there is no normal
phase. Another important result is that for $\mu\gg m$ the pseudo-Goldstone
states, which would have been massless in the SU(2$N_f$) symmetric limit,
have mass $2\mu$ \cite{KSTVZ,KST}.

Next we discuss the possibilities for Two Color Lattice QCD with adjoint quarks
\cite{HM2}.
For $N\ge2$, a diquark operator satisfying all of our ideal requirements can
always be written down: for $N=2$ it reads
\begin{equation}
qq_{\bf3}={i\over2}\left[\chi^{p\,tr}(x)\varepsilon^{pq}\chi^q(x)+
\bar\chi^p(x)\varepsilon^{pq}\bar\chi^{q\,tr}(x)\right],
\label{eq:qq3}
\end{equation}
where $p,q=1,2$ are explicit flavor indices.
In fact, just as in the fundamental case, $qq_{\bf3}$ is related to 
the adjoint chiral condensate (\ref{eq:conda}) by a global symmetry, 
in this case the
U(4) rotation given by
\begin{equation}
V={1\over2}\left(\matrix{\phantom{i}P&iP^{tr}\cr iP&\phantom{i}P^{tr}\cr}\right)
\;\;\;\mbox{with}
\;\;\;P=\left(\matrix{1&-1\cr1&\phantom{-}1\cr}\right).
\end{equation}
Condensation of $qq_{\bf3}$ breaks U(1)$_B$ but leaves unbroken an SU(2)
of isospin. Since there are diquark states among the 6 Goldstones 
expected from the breaking U(4) $\to$ Sp(4), we expect the usual 
scenario to apply, with a transition to a superfluid phase at $\mu_o=m_\pi/2$.

For $N=1$ the Exclusion Principle prevents us from writing a diquark
operator satisfying all the requirements, since
$\chi^{tr}(x)\chi(x)\equiv0$. We have considered two possibilities.
Firstly, a non-local operator which is gauge invariant and scalar under
(\ref{eq:lattpar}) is
\begin{equation}
qq_{\bf3}^\prime={1\over16}\sum_{\pm\mu}\eta_\mu(x)(-1)^{x_\mu}\left[
\chi^{tr}(x)U_\mu(x)\chi(x+\hat\mu)-\bar\chi(x)U_\mu(x)\bar\chi^{tr}(x+\hat\mu)
\right].
\label{eq:qq3p}
\end{equation}
The $(-1)^{x_\mu}$ factor ensures that $qq_{\bf3}^\prime$ is antisymmetric with
respect to spatial exchange of $\chi$ fields. In terms of the $X,\bar X$ fields,
\begin{equation}
qq_{\bf3}^\prime={1\over8}\sum_{\pm\mu}\eta_\mu(x)(-1)^{x_\mu}
\bar X_e(x)U_\mu(x)\left(\matrix{&-1\cr1&\cr}\right)X_o(x+\hat\mu).
\end{equation}
In the $m,\mu\to0$ limit the global symmetries left unbroken by 
$\langle qq_{\bf3}^\prime\rangle\not=0$ are a U(1)$_\varepsilon$ symmetry,
\begin{equation}
X\mapsto e^{-i\alpha}X\;\;\;;\;\;\;\bar X\mapsto\bar Xe^{i\alpha},
\end{equation}
which is broken when $m\not=0$, and an O(2) symmetry rotating 
$\chi$ into $\bar\chi$:
\begin{equation}
X\mapsto QX\;\;\;;\;\;\;\bar X\mapsto\bar XQ^{tr}\;\;\;;\;\;\;
Q=\left(\matrix{\phantom{-}\cos\theta&\sin\theta\cr-\sin\theta&\cos\theta\cr}
\right).
\end{equation}
The pattern of symmetry breaking is thus
U(2)$\to$U(1)$\otimes$U(1)$_\varepsilon$, leaving two unbroken generators and
hence two Goldstone modes, which turn out to be parity even meson and diquark 
states, the diquark remaining massless once $m,\mu>0$. 
Since for $m\not=0$ there are no exact Goldstones in the low density phase where
chiral symmetry is broken,
we once again predict a phase separation between a low density phase and a 
high density superfluid. Recalling, however, that the only Goldstone 
at low density is a $q\bar q$ meson,
we are unable to apply the effective theory arguments of \cite{KSTVZ,KST},
and in this case do not expect $\mu_o=m_\pi/2$.

Secondly, consider an operator which is local but not gauge-invariant:
\begin{equation}
qq_{sc}^i=
{1\over2}\left[\chi^{tr}(x)t^i\chi(x)+\bar\chi(x)t^i\bar\chi^{tr}(x)\right],
\end{equation}
whose consistency with the Exclusion Principle is now due to the antisymmetry of
the generators $t^i$. Under a gauge transformation, $qq_{sc}^i$ transforms
in the adjoint representation of SU(2),
\begin{equation}
qq_{sc}^i\mapsto O_{ij}qq_{sc}^j,
\label{eq:qqsc}
\end{equation}
which follows from the property $U^{-1}\lambda^iU=O_{ij}\lambda^j$,
true for arbitrary representations generated by $\lambda^i$.
Therefore $qq_{sc}^i$ acts like an
adjoint Higgs field, and its condensation breaks the SU(2) color group to U(1),
as in the Georgi-Glashow model of electroweak physics \cite{GG}.
As the subscript implies, this is therefore a superconducting solution; 
ironically, in this case the superconducting phase is characterised
(at least in perturbation theory)
by a {\sl massless\/} photon. Once again, because of the change in the number
of massless particles between the low density confined phase and the high
density  superconducting phase, a true phase 
separation is expected\footnote{Note that in the 2+1 dimensional SU(2) 
adjoint Higgs model, there appears to be 
no phase separation, since in principle
the photon can acquire a mass via non-perturbative effects; it
appears extremely light in the ``Higgs''
phase, however \cite{HPST}.}, which is 
consistent with general properties of gauge theories
with Higgs fields in the adjoint representation \cite{FS}, and in possible
contrast with the `color-flavor locked' state anticipated in three-flavor
QCD \cite{ARW2}, where continuity between high and low density phases has been
postulated \cite{SW}.

\subsection{The Sign of the Determinant}

As discussed in subsection \ref{subs:form}, it is straightforward to show that
the path integral measure, proportional to $\mbox{det}M$, 
is real. To determine whether it is positive definite requires
a more subtle argument. Let us first make some general observations:
\begin{itemize}

\item
Lemma 1: let $M$ be any diagonalisable operator and $K$ be the complex 
conjugation operator. If there exists a unitary operator $T$ such that
$[KT,M]=0$, then $\mbox{det}M$ is real.

\item
Proof: let $\psi$ be an eigenvector of $M$ with eigenvalue $\lambda$; then
$\tilde\psi=KT\psi$ is an eigenvector with eigenvalue $\lambda^*$:
\begin{eqnarray}
M\psi=\lambda\psi\;\Rightarrow\;KTM\psi=KT\lambda\psi\;
\Rightarrow\;MKT\psi=\lambda^*KT\psi\;\Rightarrow\;
M\tilde\psi=\lambda^*\tilde\psi\nonumber;
\end{eqnarray}
hence $\mbox{det}M$ is real, since the product over eigenvalues can be organised
in complex conjugate pairs.
\end{itemize}
Now this does not imply that $\mbox{det}M$ is positive, since for 
real $\lambda$, $\tilde\psi$ might be proportional to $\psi$, and hence
we cannot exclude the possible existence of non-degenerate real eigenvalues
which may include an odd number of negative ones. To prove that $\psi$ and
$\tilde\psi$ are linearly independent, we need $KT$ to have another property.
\begin{itemize}

\item
Lemma 2: if $(KT)^2=-1$, $\psi$ and $\tilde\psi$ are linearly independent.

\item
Proof:
\begin{eqnarray}
\langle\psi\vert\tilde\psi\rangle=
\langle\psi\vert KT\psi\rangle=\langle T\psi\vert
TKT\psi\rangle=\langle(KT)^2\psi\vert
KT\psi\rangle=-\langle\psi\vert\tilde\psi\rangle=0\nonumber,
\end{eqnarray}
where we have used $\langle\phi\vert\psi\rangle=\langle K\psi\vert
K\phi\rangle$. 
\end{itemize}
Hence if $[KT,M]=0$ and $(KT)^2=-1$, all real eigenvalues are doubly 
degenerate, and $\mbox{det}M>0$.

The square of the anti-unitary operator
$KT$ determines the Dyson index $\beta$ of $M$, the case $(KT)^2=1$ implying
$\beta=1$ and $(KT)^2=-1$ implying $\beta=4$ \cite{KSTVZ}. Physical QCD
with three colors and fundamental quarks
has no operator equivalent to $KT$, corresponding to $\beta=2$.
Now let us identify the operator $T$ for both continuum and
for staggered lattice fermions. In the continuum, 
\begin{equation}
M=(\partial_\nu+ig\lambda^iA_\nu^i)\gamma_\nu+\mu\gamma_0+m,
\end{equation}
with $A_\nu^i$ the gauge potential and $\lambda^i$ the generator appropriate 
to the quark representation. We find
\begin{eqnarray}\label{eq:contT}
\mbox{fundamental:}\;\;\;T&=&C\gamma_5\otimes\tau_2,\;\;(KT)^2=1\;\;\Rightarrow
\beta=1;\nonumber\\
\mbox{adjoint:}\;\;\;T&=&C\gamma_5\otimes\One,\;\;(KT)^2=-1
\;\;\Rightarrow\beta=4.
\end{eqnarray}
Here, $C$ is the charge conjugation matrix defined by its operation on
Euclidean hermitian $\gamma$-matrices:
$C\gamma_\mu C^{-1}=-\gamma_\mu^*$. 
For staggered lattice fermions $M$ is given by (\ref{eq:s0},\ref{eq:s1}), and
\begin{eqnarray}\label{eq:lattT}
\mbox{fundamental:}\;\;\;T&=&\tau_2,\;\;(KT)^2=-1\;\;\Rightarrow\beta=4;
\nonumber\\
\mbox{adjoint:}\;\;\;T&=&\One,\;\;(KT)^2=1\;\;\Rightarrow\beta=1.
\end{eqnarray}
Once again, we see the r\^oles of fundamental and adjoint representations
reversed for staggered lattice fermions. 
We thus have a proof that the
functional integral measure is positive definite for continuum adjoint 
quarks and fundamental lattice quarks. There is no such proof
for continuum fundamental quarks and lattice adjoint quarks, and as we
shall demonstrate in section \ref{subs:hmc}, there are indeed isolated
real eigenvalues and hence a sign problem for
the adjoint lattice model at large chemical potential $\mu$.

\subsection{The Continuum Limit}
\label{subs:cont}

In this paper our viewpoint will be to study the lattice model in its own right,
that is as a strongly coupled theory with the potential to show superfluid or
superconducting properties at high density, and not to attempt 
either continuum or chiral limits. We have therefore focussed on the global
symmetries appropriate to staggered fermions. This should not obscure the fact,
however, that there are interesting issues related to the continuum limit
whose resolution is still not clear. The model with $N$ flavors
of staggered fermion should correspond to a continuum theory with $N_f=4N$
physical quark flavors. For the model with fundamental quarks, therefore,
we expect the U(2$N$) global symmetry to enlarge to SU(8$N$), which is 
spontaneously broken by a chiral condensate to Sp(8$N$): for the adjoint
model the corresponding pattern is SU(8$N)\to\mbox{O}(8N)$. Does the 
r\^ole reversal of fundamental and adjoint representations cease at some point?
Since the behaviour of both is qualitatively 
similar in the continuum formulation, 
this is plausible \cite{KSTVZ}. On the other hand, we have argued that the 
lattice model with $N=1$ staggered flavor exhibits a distinct
behaviour, with no premature onset transition, and the possibility of a
superconducting condensate. If this were the case, 
would the superconducting phase
survive the continuum limit? Above we also argued that the sign of
the fermion determinant $\mbox{det}M$ is not positive in all cases, implying
that it is important whether $N$ is even or odd, and once again
continuum and lattice models have distinct behaviour.

Let us outline a possible route to resolving these issues. First consider
the effect of extending the continuum operator $T$ of eqn.~(\ref{eq:contT})
by including the effects of a four-index flavor structure. One can then consider
an operator
\begin{equation}
T^\prime=T\otimes\Gamma
\end{equation}
where $\Gamma$ is a $4\times4$ matrix acting on flavor; the case
$\Gamma=C\gamma_5$
has the property that $(KT^\prime)^2=-(KT)^2$, which would render 
the fermion determinant for 4 fundamental continuum flavors positive
(as must clearly be the case).
Next consider the form of the lattice $T$ operator (\ref{eq:lattT})
in a basis of fields $q^{\alpha a},\bar q^{\alpha a}$
carrying both a spinor index
$\alpha$ and a flavor index $a$ (implicit
in the staggered fermion approach), each taking values from 1 to 4 
\cite{Kluberg}. The appropriate structure is
\begin{equation}
T_q=C\gamma_5\otimes C\gamma_5\otimes\left\{\matrix{\tau_2\cr\One\cr}\right\},
\end{equation}
where the first $C\gamma_5$ acts on the spinor index, the second on 
flavor, and the final $\tau_2$ ($\One$) acts on color in the case of
fundamental (adjoint) quarks. At non-zero lattice spacing, due
to a term which is formally $O(a)$ when $M$ is expressed in the 
$q,\bar q $ basis \cite{Kluberg},
this is
the only exact symmetry of the form (\ref{eq:contT},\ref{eq:lattT}); as
we saw, $(KT_q)^2=1$ for adjoint quarks, implying a possible sign problem.
However, in the continuum limit we expect flavor symmetry to be restored, 
implying that $T^\prime_q$ with the second $C\gamma_5$
substituted by arbitrary $\Gamma$ will also define
symmetries of $M$. The existence of merely one $T^\prime_q$ with
$(KT^\prime_q)^2=-1$ (eg.\ $\Gamma=\One$) suffices to prove $\mbox{det}M$
positive due to degeneracy of real eigenvalues; we expect as a consequence
an enlargement of the Goldstone manifold to recover the continuum symmetry
prediction. Away from the continuum limit, these extra baryonic Goldstone modes
remain massive due to lattice artifacts.

We can also examine the diquark
condensates postulated in subsection \ref{subs:diquark} in terms of 
$q,\bar q$ 
\cite{HM}.
In this representation the local diquark operators (\ref{eq:qq2},
\ref{eq:qq3}, \ref{eq:qqsc})
take the form
\begin{equation}
qq={1\over2}\left[
q^{tr}(C\gamma_5\otimes C\gamma_5\otimes T)q+
\bar q(C\gamma_5\otimes C\gamma_5\otimes T)\bar q^{tr}\right],
\end{equation}
where the first operator in the tensor product in each term acts on spinor
indices, the second on flavor, the third 
on gauge indices (or explicit flavor indices in the case of (\ref{eq:qq3})),
and all three are antisymmetric matrices.
In the same basis the parity transformation
(\ref{eq:lattpar}) reads
\begin{equation}
q(x)\mapsto(\gamma_0\otimes\gamma_5\otimes\One)q(x^\prime)\;\;\;;\;\;\;
\bar q(x)\mapsto\bar q(x^\prime)(\gamma_0\otimes\gamma_5\otimes\One).
\label{eq:contpar}
\end{equation}
We thus confirm that the first $C\gamma_5$ is consistent with the local
condensates being spacetime scalars, the second $C\gamma_5$ implying that the
condensates transform in the antisymmetric tensor representation
of the continuum flavor SU(4)
symmetry expected away from the chiral limit, which has dimension {\bf6}.
The situation is different for the non-local condensate
(\ref{eq:qq3p}), which reads
\begin{equation}
qq_{\bf3}^\prime=\sum_\mu q^{tr}(C\otimes C\gamma_\mu^*\otimes\One)q
-\bar q(C\otimes C\gamma_\mu^*\otimes\One)\bar q^{tr}.
\end{equation}
Invariance of $qq_{\bf3}^\prime$ under parity depends on the non-trivial action
of (\ref{eq:contpar}) on flavor, and in fact the spacetime structure
of the operator implies that this condensate is {\sl pseudoscalar\/}.
The symmetry of $qq_{\bf3}^\prime$ may be checked noting that
$C\gamma_\mu^*$ is a symmetric matrix, in turn implying that this 
condensate transforms as a symmetric {\bf10} of flavor SU(4).
Without further detailed dynamical input, it is difficult to proceed; it may
well prove that the route to the continuum is
complicated, with several distinct phases found as the parameters
$\beta$, $m$ and $\mu$ are tuned.

Finally, we note that because adjoint sources screen color
four times more effectively
than fundamental ones, asymptotic freedom is lost in Two Color QCD with $N_f$
massless adjoint quarks for $N_f=4N>11/4$, ie.\ even for $N=1$. 
Therefore in the chiral limit even the
location of the continuum limit is {\it a priori\/} unknown; possible behaviours
include a continuum limit described by a non-perturbative renormalisation
group fixed point at some finite 
value of the gauge coupling
$\beta_c$, or no interacting continuum limit existing at all. In 
section \ref{sec:results} 
we will demonstrate that these considerations are irrelevant at the
values of $\beta$ and $m$ considered in our work.

\subsection{Summary}

In this section we have reviewed the formulation and symmetries of Two Color
lattice QCD at length, and compared and contrasted it with corresponding 
continuum
models. We conclude that the version of Two Color QCD
which is most `QCD-like' with respect to non-zero chemical potential
$\mu$ and hence worthy of study, is
that with $N=1$ flavor of staggered lattice fermion
in the adjoint representation, for the following reasons:

\begin{itemize}

\item 
There are no Goldstone diquarks in the spectrum at zero density, 
and hence 
no reason to expect a premature onset transition at $\mu_o\simeq m_\pi/2$.

\item
The spectrum in the confined phase should contain gauge invariant 
fermionic bound states, either
of a quark and a gluon, or of an 
odd number of quarks. Hence there is the possibility of a nuclear liquid 
phase, and the formation of a Fermi surface,
before the restoration of chiral symmetry expected as $\mu$ is increased.

\item
The path integral measure is real but not positive definite for any odd
$N$; that of QCD
is complex. It may prove possible to expose the 
importance of configurations with non-positive definite 
$\mbox{det}M$
in the path integral with $\mu\not=0$.

\item
There is a possibility of a gauge-variant diquark condensation at high
densities, leading to color superconductivity. No other lattice model
capable of being studied with $\mu\not=0$ appears to share this feature
\cite{HKLM,HM,MH}.

\end{itemize}
It is conceivable that some or all of these factors are intimately
linked.

\section{Simulation Algorithms}
\label{sec:algo}

We have studied Two Color lattice QCD with adjoint 
staggered fermions
using two different simulation algorithms, the hybrid Monte Carlo (HMC)
algorithm \cite{DKPR}, and
a Two-Step Multi-Bosonic (TSMB) algorithm \cite{TWO_STEP}. 
Here we outline the two methods,
and point out some important features of each.

\subsection{The Hybrid Monte Carlo Algorithm}
\label{subs:hmc}

The HMC algorithm starts 
from an expression for the
action $S$ in terms of bosonic auxiliary pseudofermion variables $\Phi$:
\begin{equation}
S={1\over2}\sum_{x,y}\sum_{p=1}^N\Phi^{p\,tr}(x)(M^{tr}M)^{-1}_{x,y}[U,\mu]
\Phi^p(y)
-{\beta\over2}
\sum_{x,\mu<\nu}\mbox{Tr\,}U_{\mu\nu}(x),
\end{equation}
where the fermion matrix $M$ is defined in (\ref{eq:s0},\ref{eq:s1}),
$\beta$ is the gauge coupling constant, and the trace over the plaquette
is taken in the fundamental representation. In contrast to QCD, the $\Phi$
may be taken real, since $M$ is real. Gaussian integration over
$\Phi$, which is convergent since the eigenvalues of $(M^{tr}M)^{-1}$ are
real and positive, yields a factor proportional to
\begin{equation}
\exp(-S_{eff})=(\sqrt{\mbox{det}M^{tr}M})^N=\vert\mbox{det}M\vert^N.
\end{equation}
This coincides with the correct functional measure $\mbox{det}^NM$ if
$\mbox{det}M$ is positive. In this case HMC correctly simulates $N=1$ flavor
of staggered lattice fermion. Note that for $\mu\not=0$, the usual trick
of evaluating $(M^{tr}M)^{-1}$ on just even lattice sites, thus reducing 
the effective number of fermion degrees of freedom by a further factor of two, 
is not available since the chemical potential spoils the anti-hermitian
form of $D$, the off-diagonal part of $M$.

It is instructive to analyse the performance of the algorithm by considering
the eigenvalue spectrum of $M$.
Because $D$ only connects even sites with odd and vice-versa, 
if $\psi(x)$ is an
eigenvector with eigenvalue $\lambda=m+\kappa$, 
then so is $\varepsilon(x)\psi(x)$, with
eigenvalue $\lambda=m-\kappa$. 
Because $D$ is a real matrix, if $\psi$ is an
eigenvector with eigenvalue $\lambda$ then so is $\psi^*$ with eigenvalue
$\lambda^*$.
Now for $\mu=0$, the matrix $D$ is anti-hermitian, implying its eigenvalues are
pure imaginary; 
we deduce that in this case the 
spectrum of $M$ lies along
the line $\lambda=m+iz$, with $z$ real.
For $\mu\not=0$, the symmetries between $\lambda$, $\lambda^*$
and $2m-\lambda$ persist, but now the spectrum swells out to occupy a roughly
elliptical region of the complex plane, whose horizontal dimension grows with
increasing $\mu$.
\begin{figure}[htb]
\psdraft
\centerline{
\setlength\epsfxsize{400pt}
\epsfbox{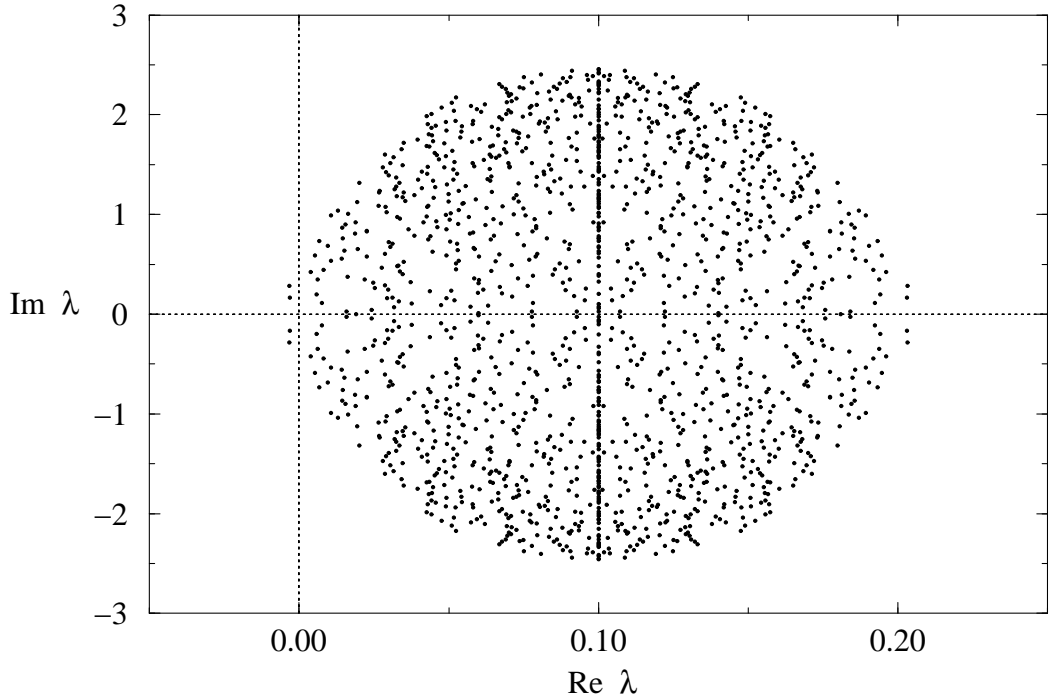
}}
\psfull
\caption{Eigenvalue spectrum from a $4^3\times8$ lattice with $\beta=2.0$,
$m=0.1$ and $\mu=0.35$. Note the mismatch in scale between real and imaginary
axes.
\label{fig:whole_spec}}
\end{figure}
The spectrum from a representative configuration is shown in 
Fig.~\ref{fig:whole_spec}. A feature of interest is that even once the 
spectrum has swelled out, an appreciable fraction of the eigenvalues remain
on the line $\lambda=m+iz$; this contrasts with behaviour observed in
three-color QCD, where all eigenvalues leave the line once $\mu\not=0$
\cite{Barbour}, but is similar to the spectrum of the random matrix Dirac
operator with $\mu\not=0$ and Dyson index $\beta=1$ (corresponding to 
a matrix $D$ with off-diagonal elements
chosen from a chiral Gaussian Orthogonal Ensemble) \cite{HOV}. 
Secondly, for this particular value of
$\mu=0.35$ the spectrum has broadened sufficiently for the extremal eigenvalues
to have negative real parts.

\begin{figure}[htb]
\psdraft
\centerline{
\setlength\epsfxsize{400pt}
\epsfbox{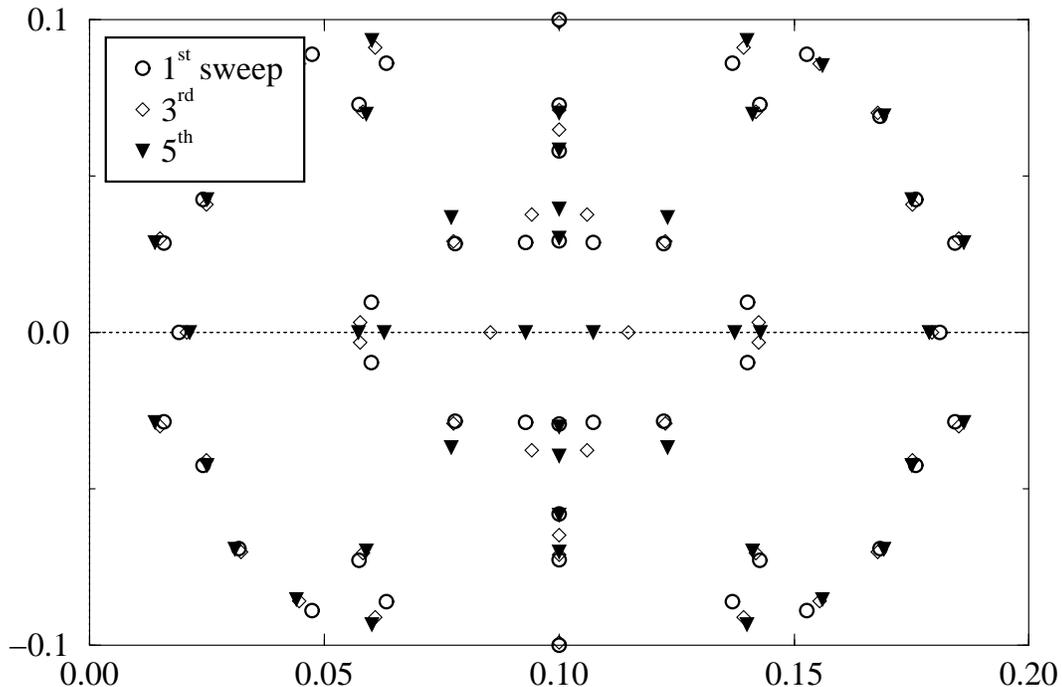
}}
\psfull
\caption{Spectral flow under HMC evolution.
\label{fig:flow}}
\end{figure}
The HMC algorithm works by evolving the $\{U\}$ 
fields in fictitious time $\tau$
for a period called a trajectory,
whereupon the resulting
configuration is accepted or rejected via a Metropolis step; 
the acceptance probability
is related to how well the pseudo-Hamiltonian flow conserves energy.
At the start of each trajectory the pseudofermion and conjugate momentum 
fields are refreshed from a Gaussian heat-bath, which ensures ergodicity in 
most cases. The trajectory length may be held constant or picked at random,
but is usually chosen to be $O(1)$. We find, however, that once 
the eigenvalue occupation region has swelled to include the origin,
the resulting presence of very small eigenvalues causes numerical
instabilities, and necessitates smaller trajectory lengths --- the data of
Fig.~\ref{fig:whole_spec} were generated using an average trajectory length of
0.18.
An HMC update step acts on the configuration as a whole,
and high acceptance is maintained by making a sequence of small changes: HMC is
thus a {\sl small step-size\/} algorithm. In
Fig.~\ref{fig:flow} we show the spectral flow under HMC evolution
in a small region close to the origin, starting from the configuration of
Fig.~\ref{fig:whole_spec} and evolving for five very short
trajectories of length 0.003. For
clarity only the eigenvalues from the first, third and fifth steps are shown.
Note that there are a few eigenvalues which are real: a pair close to 
0.1$\pm$0.08 on the first configuration, which increases to eight by the fifth.
Close inspection reveals that between the first and third configurations a pair 
of eigenvalues jumps from the line $\lambda=m+iz$ at approximately
$0.1\pm0.03i$ to the real axis at
$\simeq0.1\pm0.02$ --- such processes change the number of real eigenvalues
by $\pm2$, and the number on the line $\lambda=m+iz$ by $\mp2$. 
Between the third and fifth configurations, two pairs of conjugate
eigenvalues coalesce and then move independently along the real axis
at $\simeq0.1\pm0.04$ --- this process changes the number of real eigenvalues by
$\pm4$. Finally between the third and fifth configurations we can also see
pairs of eigenvalues coalesce and then move independently along $m+iz$
at $\simeq0.1\pm0.04i$, changing the number of eigenvalues on this line by
$\pm4$. All other eigenvalues in the region, at generic points in the complex
plane, merely exhibit small fluctuations under HMC
evolution. The symmetries between $\lambda$, 
$\lambda^*$ and $2m-\lambda$ are maintained at all times.

Fig.~\ref{fig:flow} demonstrates unambiguously the existence of non-degenerate
real eigenvalues of $\lambda$, and the fact that there is always an even number
of them. There is no obstruction in principle to there being an
odd number of negative real eigenvalues, resulting in $\mbox{det}M<0$. However,
since 
the processes discussed in the previous paragraph always create real eigenvalues
in pairs at the same point, 
the only route to $\mbox{det}M<0$ can be if an isolated eigenvalue
moves along the real axis and through the origin, forcing 
$\mbox{det}M$ to evolve smoothly through zero. In the neighbourhood of such a 
point the effective action $S_{eff}$ diverges, implying a strong repulsive 
force in the Hamiltonian flow, and hence a large kinetic barrier 
to changing the 
determinant's sign. This barrier is a feature of any model
in which $\mbox{det}M$ is real; in QCD, which has $\mbox{det}M$ complex,
it should be straightforward to change the sign by moving through a sequence 
of configurations in which the phase of the determinant gradually increases
from 0 to $\pi$.

\begin{figure}[htb]
\psdraft
\centerline{
\setlength\epsfxsize{400pt}
\epsfbox{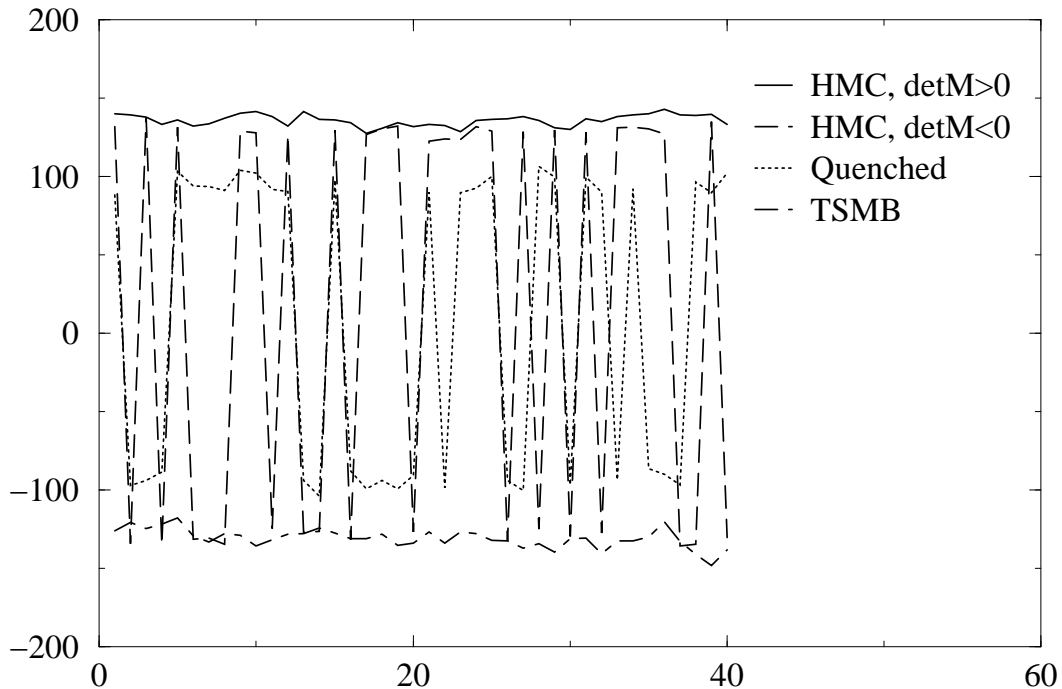
}}
\psfull
\caption{Evolution of $\vert\ln\mbox{det}M\vert\times\mbox{sign}(
\mbox{det}M)$ using
HMC, quenched and TSMB updates on a $4^3\times8$ lattice at $\beta=2.3$, $m=0.1$
and $\mu=0.6$.
\label{fig:detsign}}
\end{figure}

In Fig.~\ref{fig:detsign} we show evolution of $\ln\mbox{det}M$ for 40
representative configurations, each separated by 50 trajectories of average
length 0.05. The simulation parameters have been chosen such that
the left-hand edge of the spectrum extends well beyond the imaginary axis, 
so that 
we expect a non-vanishing probability of obtaining negative real eigenvalues.
We see on inspection of the first, solid curve that HMC evolution does not
appear to change the sign of $\mbox{det}M$. This could in principle be because
the volume of configuration space with an odd number of real negative
eigenvalues is very small; we can eliminate this possibility by considering 
a comparable sequence of quenched updates (dotted curve). In this case the sign
of $\mbox{det}M$ is seen to change frequently, confirming that such
configurations exist and are relatively easy to find. Moreover, if one of the
configurations with $\mbox{det}M<0$ is then reequilibrated with HMC and allowed
to evolve, we see that once again the sign remains stable and negative
(dot-dashed curve); this behaviour is observed to persist for both signs 
for approaching 20000 trajectories. We reach two important conclusions:

\vfill\eject
\begin{itemize}

\item
There are regions of parameter space with $\mu\not=0$ for which 
$\mbox{det}M$ can take negative values, ie.\ there is
a sign problem.

\item
The kinetic barrier at the origin prevents
the HMC algorithm from changing the sign of $\mbox{det}M$, and therefore 
from exploring the whole of the system's configuration space: in other words,
the HMC algorithm is not {\sl ergodic\/} in this region of parameter space.

\end{itemize}
The sign problem is well-known 
\cite{sign} 
and can be addressed, at least in principle, by including
the sign of $\mbox{det}M$ with the observable as in (\ref{eq:obs});
this may be expected to be
effective provided the average sign is significantly different from
zero.
The problem with ergodicity is less well-known --- it can be anticipated in any
model where $M$ is real but its eigenvalues $\lambda$ complex;
another example which would be interesting to study is the lattice
Gross--Neveu
model with discrete Z$_2$ chiral symmetry \cite{HKK}. Note that it remains a
problem of principle for even $N$, since
we can still classify a configuration by the number of negative real
eigenvalues of $M$. It is an interesting open question whether or not the
absence of transitions between odd and even sectors is of practical importance.

Despite its problems, we have made an extensive study of Two Color QCD using 
the HMC algorithm, and the results will be surveyed in section
\ref{sec:results}. 
We next turn to an algorithm which has the potential to overcome the ergodicity
problem.

\subsection{The Two-Step Multi-Bosonic Algorithm}
\label{subs:tsmb}

For the Two-Step Multi-Bosonic algorithm \cite{TWO_STEP} we 
also need a hermitian fermion matrix.
First one might consider
\begin{equation} \label{Q_tilde}
\tilde{Q}(\mu) \equiv \varepsilon M(\mu) =
\tilde{Q}(-\mu)^\dagger,
\end{equation}
but for non-zero chemical potential this is still not hermitian.
Hermiticity may be achieved by doubling the matrix size:
\begin{equation} \label{Q_hat}
\hat{Q}(\mu) \equiv \left( 
\begin{array}{cc}  0               &  \tilde{Q}(-\mu)   \\ 
                   \tilde{Q}(\mu)  &  0   \end{array}   \right)
= \hat{Q}(\mu)^\dagger \ .
\end{equation}
Following from
\begin{eqnarray} \label{Q_hat_square}
\hat{Q}(\mu)^2 & = & \left( 
\begin{array}{cc}  \tilde{Q}(-\mu) \tilde{Q}(\mu)  &  0  \\ 
                   0   &   \tilde{Q}(\mu)\tilde{Q}(-\mu)   \end{array}   \right)
\nonumber \\[0.5em]
& = &  \left( 
\begin{array}{cc}  M(\mu)^\dagger M(\mu)  &  0  \\ 
                   0   &   M(-\mu)^\dagger M(-\mu)   \end{array}   \right)
\end{eqnarray}
and noting that
\begin{equation} \label{det_Q}
\det M(\mu) = \det M(\mu)^* = \det M(-\mu), 
\end{equation}
(where the first equality holds because $M=D+m$ is real,
the second because $D^{tr}(\mu)=-D(-\mu)$, and the spectrum of $D$
is symmetric about zero),
we deduce
\begin{equation} \label{det_Q_hat}
\det \hat{Q}(\mu)^2 = \left\{ \det M(\mu) \right\}^4 \ ,
\hspace{2em}
| \det M(\mu) | = \left\{ \det \hat{Q}(\mu)^2 \right\}^{1 \over 4} \ .
\end{equation}

 In multi-bosonic representations of the fermion determinant \cite{LUSCHER}
 polynomials are used to approximate the necessary inverse powers of
 $x \equiv \hat{Q}(\mu)^2$ over some prescribed range $x\in[\epsilon,\lambda]$:
\begin{equation} \label{inverse_quarter}
\frac{1}{x^{1 \over 4}} \simeq P(x) \ , 
\hspace{2em}
| \det M(\mu) | \simeq \frac{1}{\det P\left( \hat{Q}(\mu)^2 \right)} \ .
\end{equation}
 This shows that in the present case the same polynomial approximations
 can be used as in recent numerical simulations of supersymmetric
 Yang Mills theory \cite{DESYMUNSTER1}.

 The usual way to represent the determinant of the polynomial in
 (\ref{inverse_quarter}) is by functional integration
over complex pseudofermion boson fields.
 Since the fermion matrix is real, it is also possible to
 use a real multi-bosonic representation, just as in the HMC case.
 For this we need a different polynomial approximation:
\begin{equation} \label{inverse_half}
\frac{1}{x^{1 \over 2}} \simeq \bar{P}(x) \ , 
\hspace{2em}
| \det M(\mu) | \simeq \frac{1}
{\left\{ \det \bar{P}\left( \hat{Q}(\mu)^2 \right) \right\}^{1 \over 2}} \ .
\end{equation}
 Since the polynomial $\bar{P}(x)$ is supposed to have complex conjugate
 pairs of roots, one can decompose it, with an overall factor $r_0$, as 
\begin{equation} \label{real_polynomial}
\bar{P}(\hat Q(\mu)^2) = r_0 \prod_j[\hat{Q}(\mu)^2+r_j] 
= r_0 \prod_j [ (\hat{Q}(\mu)+\mu_j)^2+\nu_j^2 ] \ ,
\end{equation}
with $\mu_j$, $\nu_j$ real.
 In order to achieve this form one has to choose the signs of the
 square roots of two complex conjugate roots appropriately:
 $r_j = (i\mu_j+\nu_j)^2 , \ r_{j+1} = r_j^* = (-i\mu_j+\nu_j)^2$.
 The multi-bosonic representation with real pseudofermion fields is
then
\begin{equation} \label{real_representation}
\frac{1}
{\left\{ \det \bar{P}\left( \hat{Q}(\mu)^2 \right) \right\}^{1 \over 2}}
\propto \int [d\Phi] \exp \left( -\sum_{jyx} 
\Phi_y^{j\,tr} [(\hat{Q}(\mu)+\mu_j)^2+\nu_j^2]_{yx} \Phi_x^j \right) \ . 
\end{equation} 
 The polynomial orders for a sufficiently good approximation in
 (\ref{inverse_half}) are typically somewhat higher than in
 (\ref{inverse_quarter}) but the use of real fields has the advantage
 of taking half the storage and roughly half the arithmetic.

In the TSMB algorithm the polynomials in 
eqs. (\ref{inverse_quarter}) and (\ref{inverse_half})
are obtained by a product of lower order polynomials.
The multi-bosonic representation is taken for the first factor 
$P^{(1)}_{n_1}(x)$ with a relatively low order $n_1$.
This diminishes the storage requirements and improves autocorrelations.
A better approximation of the fermion determinant is achieved by a
second polynomial factor $P^{(2)}_{n_2}(x)$ of order $n_2$.
In the gauge field update the effect of $P^{(2)}$ is taken into
account stochastically.
Another auxiliary polynomial $P^{(3)}_{n_3}(x)$ is also needed in this
stochastic correction.
The final precision in the approximation is achieved by reweighting
the gauge configurations when evaluating expectation values.
There a fourth polynomial $P^{(4)}_{n_4}(x)$ is used.
For appropriate algorithms to obtain the necessary optimized polynomial
approximations see \cite{IM2}.
A detailed description of the TSMB algorithm can be found in
\cite{DESYMUNSTER2}.

There are two reasons why the TSMB algorithm can overcome the 
ergodicity problem related to the zero of the fermion determinant
discussed in section \ref{subs:hmc}.
First, the gauge field updates are performed by some {\sl large step-size\/}
algorithm, in our case the Metropolis algorithm.
Second, the imperfect approximations near the zero of the determinant
open a ``hole'' where the tunnelling between the sectors
of differing determinant sign is facilitated.
The small error in the evaluation of expectation values
due to the imperfection of the approximation can be
removed by the reweighting step \cite{FJ}.
In Fig.~\ref{fig:detsign} we show as a dashed line the evolution of
$\vert\ln\mbox{det}M\vert\times\mbox{sign(det}M)$
(for the uncorrected $M$), using a TSMB algorithm and similar simulation 
parameters to the other algorithms in the figure. The algorithm
appears to be capable of changing the sign of the determinant 
if anything slightly more effectively than the quenched updates.

A possible procedure to tune the parameters of the TSMB algorithm is as
follows: since the HMC algorithm is also available, one can
determine the smallest ($\lambda_{min}$) and largest ($\lambda_{max}$)
eigenvalues of $\hat{Q}^2$ on typical gauge configurations from a
HMC run. 
The lower and upper limits of the approximation interval 
$[\epsilon,\lambda]$ can be chosen as $\epsilon \simeq 0.5\lambda_{min}$
and $\lambda \simeq 1.5\lambda_{max}$.

The order of the polynomial for the noisy correction $n_2$ can be taken
to be roughly the same as the average number of iterations in the 
inversions of HMC.
The third polynomial used in the noisy correction should have an order
$n_3$ which is typically 10-30\% larger than $n_2$.
(A good test for $n_3$ is that the noisy correction should ideally
always accept an unchanged gauge configuration.
An acceptance of about 99\% is sufficient in practice.)
After fixing $n_2$ one can optimise the order of the first polynomial
by tuning the average acceptance of the noisy correction step.
An acceptance of about 60-70\% turned out to be optimal in most cases.
Note that the quality of approximation of the fermion determinant
is practically independent of $n_1$ once $n_2$ is fixed.

An interesting parameter in the optimization of the autocorrelation
is the ratio of update sweeps performed on the bosonic pseudofermion 
fields versus gauge fields.
This depends on the lattice parameters and also on the machine, code 
optimization, compiler etc.
In our case we typically found it better to choose
two or three times as many gauge 
updates as pseudofermion updates.
This differs from the experience in supersymmetric Yang-Mills theory
\cite{DESYMUNSTER2} where it is better to have relatively more boson 
field updates.

The final step is to check the approximation of the fermion determinant 
by calculating the reweighting factors on the measured gauge 
configurations.
This can be done by determining a few (say, 8 or 16) of the smallest eigenvalues
of $\hat{Q}^2$ and calculating the reweighting factor explicitly for them,
and afterwards multiply by the stochastic reweighting factor obtained 
by a large order ($n_4$) polynomial on the orthogonal subspace.
In practice the reweighting does not change the averages if the 
reweighting factors are within a few percent of unity.
In runs with large condition numbers $\lambda/\epsilon$,
up to $\lambda/\epsilon\simeq10^7$ for our parameters, this is not the 
case and the reweighting is important.
In fact, in these cases it is not optimal to try to increase $n_2$
so long as the reweighting is negligible, since the 
very high orders required would slow down the updating too 
much.
It is better to increase $n_2$ only up to the level that the reweighting
factors are typically $O(1)$.
The final precision is then achieved by reweighting and this has to be 
done typically only on (more or less) independent configurations.

Note that if HMC is not available for starting the optimisation one
can first consider a point with heavy fermions where low order 
polynomials are sufficient and gradually decrease the mass to the
point of interest.

Let us mention a technical point which turns out to be important for
dealing with large order polynomials, especially in case of a single 
precision (32 bit) computation.
The relevant variable for the polynomials is the condition number
$\lambda/\epsilon$.
The actual values of $\lambda$ and $\epsilon$ can be reached by 
rescaling.
This enables very large or very small numbers
appearing in the expansion coefficients of the polynomials to be avoided.
It turns out that choosing, for instance, $\lambda=4$ keeps these
numbers within a reasonable range.
The required rescaling factor can then be included in the recursive
evaluation of the polynomials.
In this way a single precision calculation becomes possible even
for polynomial orders $n = O(1000)$.

\section{Results}
\label{sec:results}

\subsection{Studies at $\mu=0$}

\begin{figure}[htb]
\psdraft
\centerline{
\setlength\epsfxsize{350pt}
\epsfbox{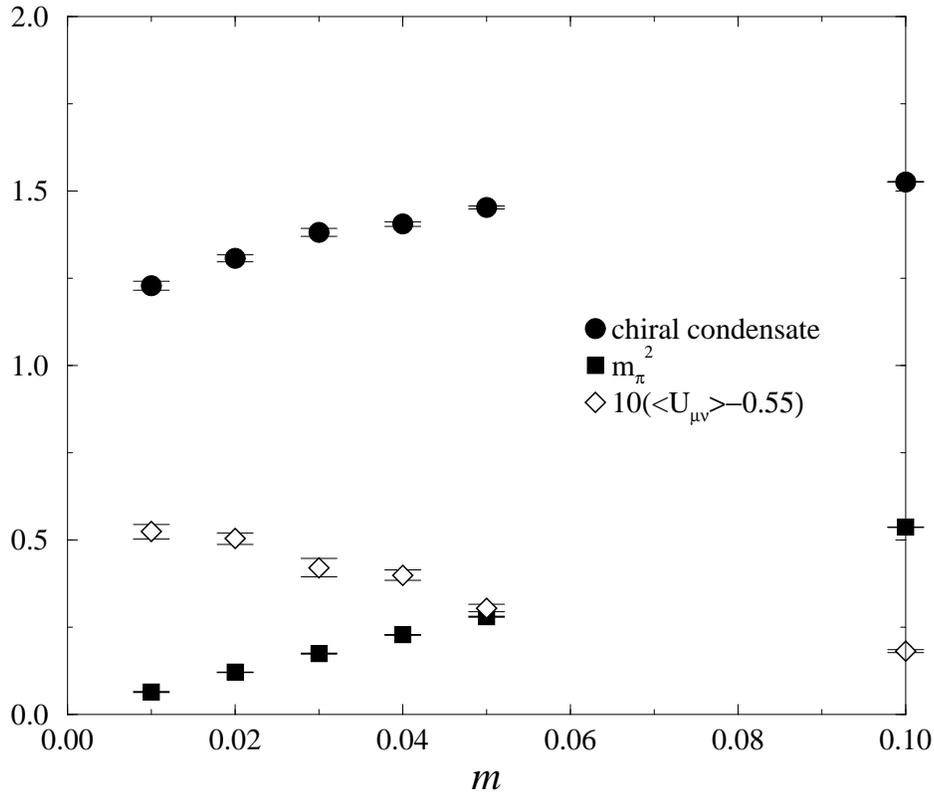
}}
\psfull
\caption{Results of HMC simulations at $\beta=2.0$ on a $4^3\times8$ lattice.
\label{fig:mu=0.0}}
\end{figure}

We have performed the bulk of our simulations in this initial study on a
$4^3\times8$ lattice with gauge coupling $\beta=2.0$ and quark masses
$m=0.1$, 0.05 and 0.01. It is important first to address the issue raised 
in subsection \ref{subs:cont}, namely whether the model exhibits confinement and
chiral symmetry breaking with these parameters at $\mu=0$, 
or whether the quarks are already 
sufficiently light to destroy asymptotic freedom and perhaps
send the theory 
into a different phase.

\begin{table}[ht]
\setlength{\tabcolsep}{1.5pc}
\caption{Results from simulations with $\mu=0$}
\label{tab:pionmu0}
\begin{tabular*}{\textwidth}{@{}l@{\extracolsep{\fill}}llll}
\hline
$m$ &  $\langle\bar\chi\chi\rangle$  & $m_\pi$   & $\Box$\\
\hline
0.10 & 1.526(1) & 0.7327(4)  & 0.5682(4)  \\
0.05 & 1.453(4) & 0.5298(16) & 0.5805(11) \\
0.04 & 1.405(6) & 0.4778(14) & 0.5899(15) \\
0.03 & 1.381(11)& 0.4181(15) & 0.5921(26) \\
0.02 & 1.307(10)& 0.3482(18) & 0.6004(16) \\
0.01 & 1.228(8) & 0.2541(25) & 0.6033(14) \\
\hline
\end{tabular*}
\end{table}
Fig.~\ref{fig:mu=0.0} shows results obtained by HMC simulation at $\mu=0$
as a function of $m$. The observables monitored, 
tabulated in Table \ref{tab:pionmu0}, are the chiral condensate 
$\langle\bar\chi\chi\rangle$ measured using a stochastic estimator, the 
pion mass $m_\pi$ estimated using a standard cosh fit to all 8 timeslices of
the pion propagator, and the average plaquette 
$\Box={1\over2}\mbox{Tr\,}\langle U_{\mu\nu}\rangle$, 
which has been rescaled for 
the convenience of the plot. The data clearly support a scenario with
$\lim_{m\to0}\langle\bar\chi\chi(m)\rangle\not=0$ and $m_\pi\propto\surd m$
in the same limit. This suggests that simulations with $m\geq0.01$ fall safely
within the regime where chiral symmetry is broken;
this suffices for our purposes, since chiral symmetry restoration 
is the main physical issue at high density. The increase in the value of 
the plaquette as $m$ is reduced shows nonetheless that 
color screening due to dynamical fermion effects is clearly observable.

\subsection{Autocorrelation Analysis}
\label{subs:auto}
\begin{figure}[htb]
\psdraft
\centerline{
\setlength\epsfxsize{400pt}
\epsfbox{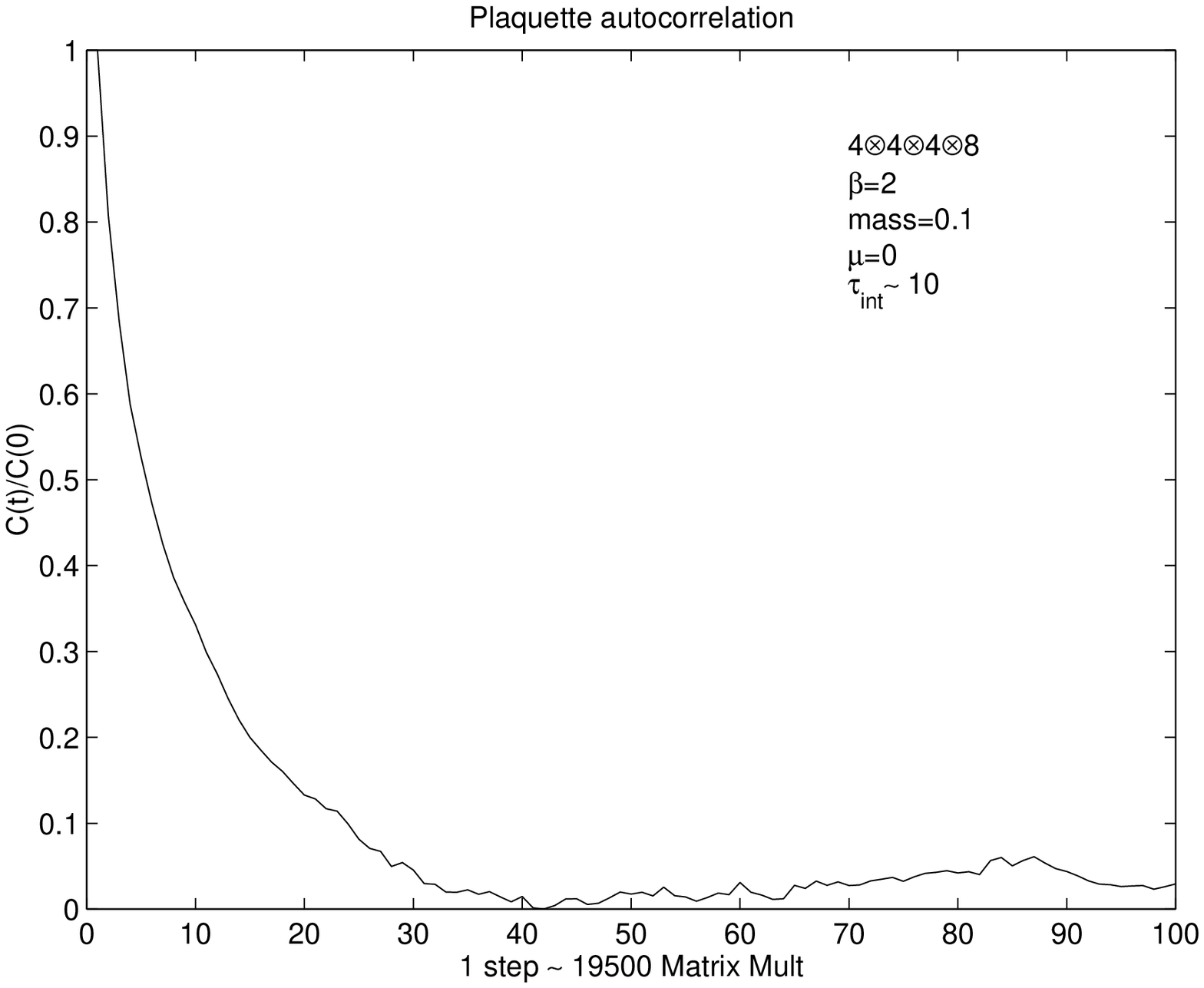
}}
\psfull
\caption{Plaquette autocorrelation function for HMC at $\mu=0.0$
\label{fig:ac_HMC_00}}
\end{figure}
\begin{figure}[htb]
\psdraft
\centerline{
\setlength\epsfxsize{400pt}
\epsfbox{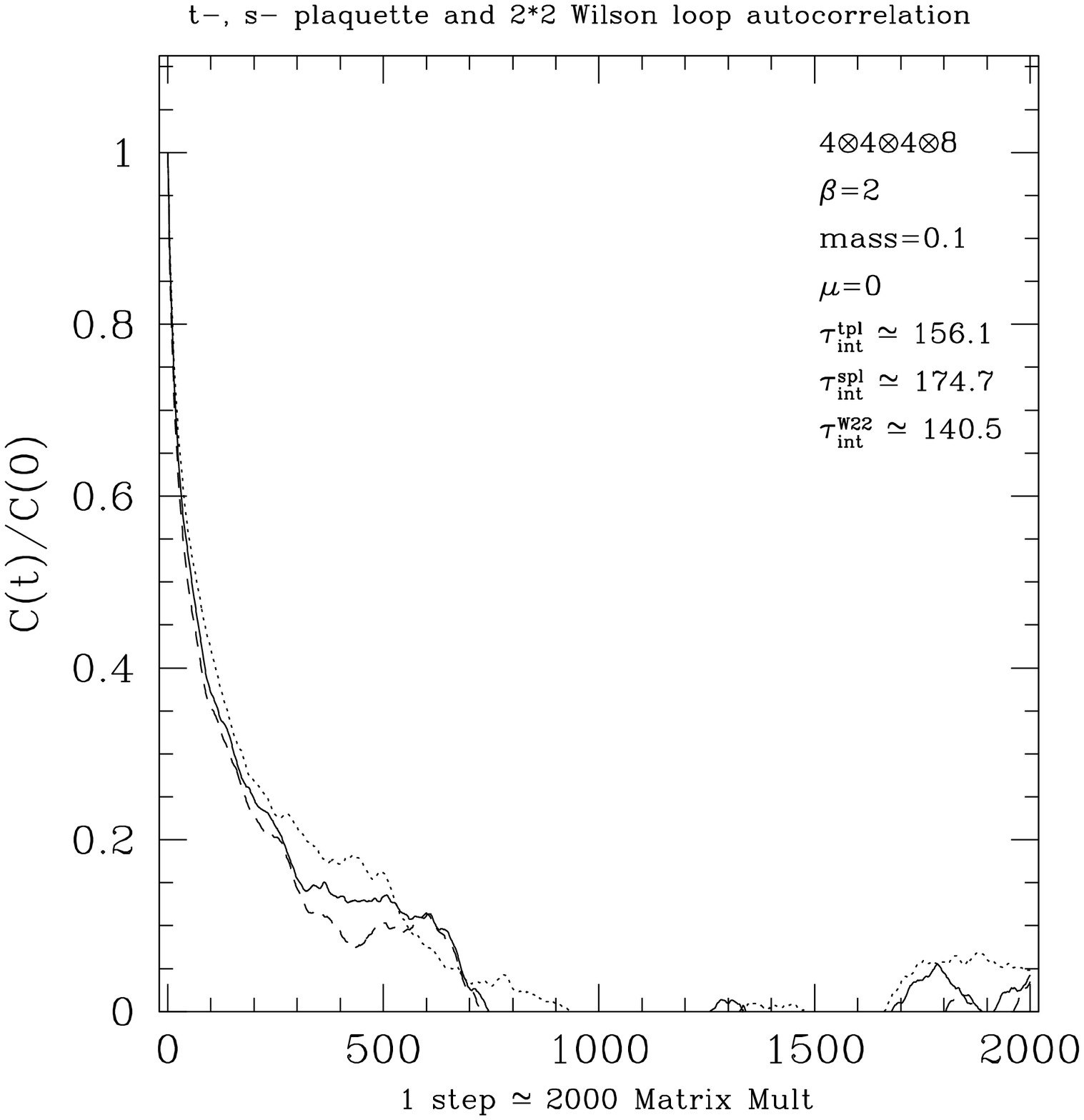
}}
\psfull
\caption{Plaquette autocorrelation function for TSMB at $\mu=0.0$
\label{fig:ac_TSMB_00}}
\end{figure}

Before exploring the phase diagram we performed some long runs
in order to determine the decorrelation time of both algorithms.
A crucial question is how the simulation effort changes for 
both HMC and TSMB as one follows the $\mu$ axis. To be able to
compare two different algorithms we need a common unit of measure
for the simulation time necessary to obtain two independent
configurations. A convenient choice is the number of matrix
multiplications (appropriately corrected with a factor that 
takes into account that TSMB spends more time in other kinds 
of operation).
For HMC the number of matrix multiplications between two
successive data takings is given by:
\begin{equation}
[({{T^{trj}}\over{d\tau}})  (I_{cg}^H) + (I_{cg}^M) ] 
\times 2 \times 2
\end{equation}
Here $T^{trj}$ is the average trajectory length 
and $d\tau$ the elementary step length for the hamiltonian 
deterministic evolution. $I_{cg}^H$ is the average 
number of conjugate gradient iterations for a single step in the 
hamiltonian evolution, and $I_{cg}^M$ is the one 
needed in the Metropolis selection step.
One factor 2 is there because each conjugate gradient iteration
implies 2 matrix multiplications, the other because
the data are printed out every 2 trajectories.

The corrected number of
matrix multiplications to perform a TSMB step is:
\begin{equation}
(n_2+n_3)\times {I_M\over2} \times F
\end{equation}
Here $I_M$ is the number of Metropolis iterations in a single TSMB step
and $F$ is the correction factor mentioned above: in practice $F$ is found
to depend on every parameter in the simulation, but has typical values
$1.2<F<2$. 

Since we found the plaquette to be the observable with 
by far the longest autocorrelation we concentrated on that
for the following analysis.
First we considered a point at $\mu=0$.
The autocorrelation function for a run with HMC is shown 
in Fig.~\ref{fig:ac_HMC_00}. We deduce an integrated
autocorrelation
time of the order of $2\times10^5$ matrix multiplications. The 
corresponding plot for a TSMB run is displayed in
Fig.~\ref{fig:ac_TSMB_00}. In this case an optimal choice of
parameters was found (following the prescription described above) to be
$n_1=24$, $n_2=90$, $n_3=110$, $I_M=12$. The integrated autocorrelation
time turns out to be about $3.2\times10^5$ matrix multiplications.

\begin{figure}[htb]
\psdraft
\centerline{
\setlength\epsfxsize{400pt}
\epsfbox{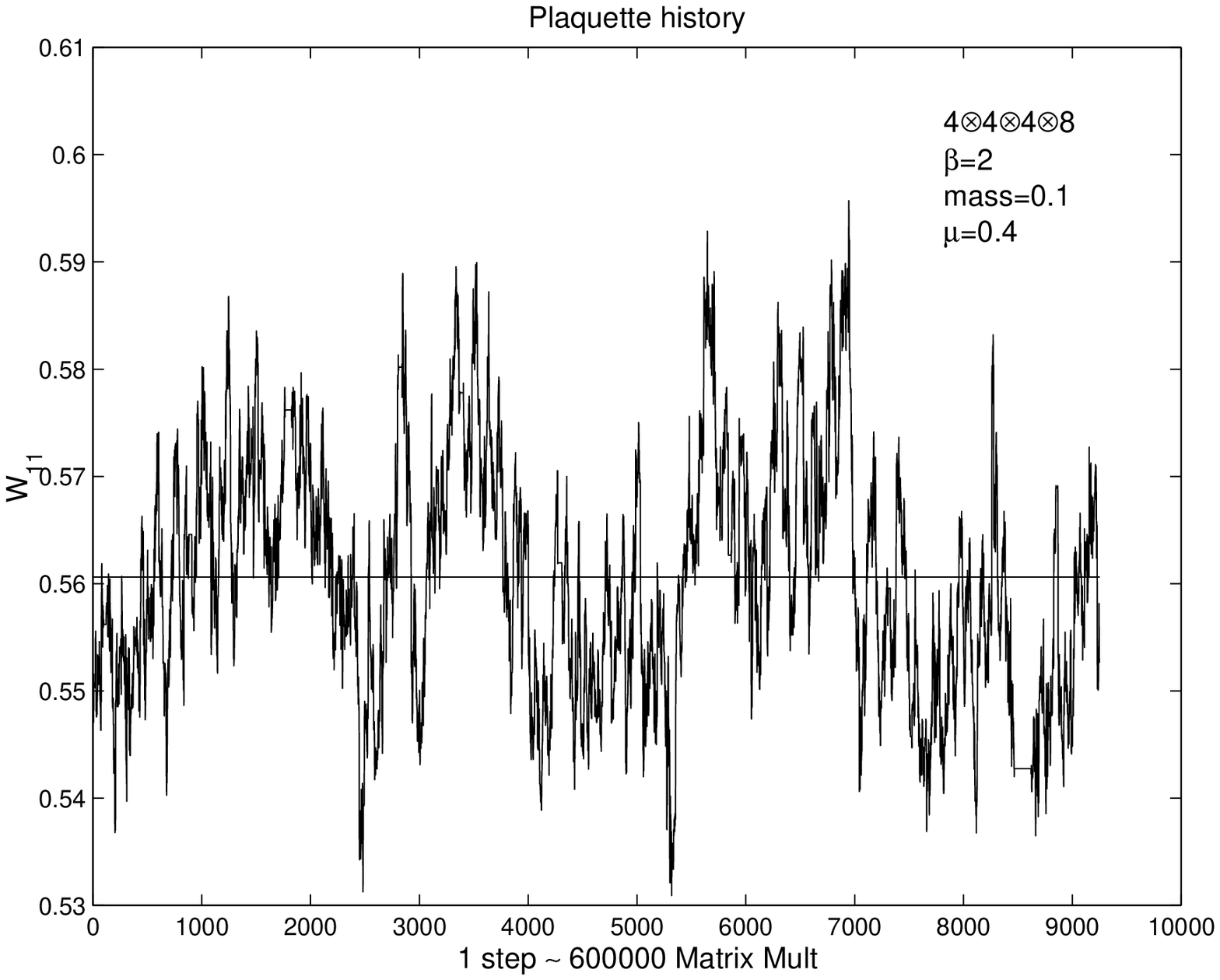
}}
\psfull
\caption{Plaquette time history for HMC at $\mu=0.4$
\label{fig:hi_HMC_04}}
\end{figure}

\begin{figure}[htb]
\psdraft
\centerline{
\setlength\epsfxsize{400pt}
\epsfbox{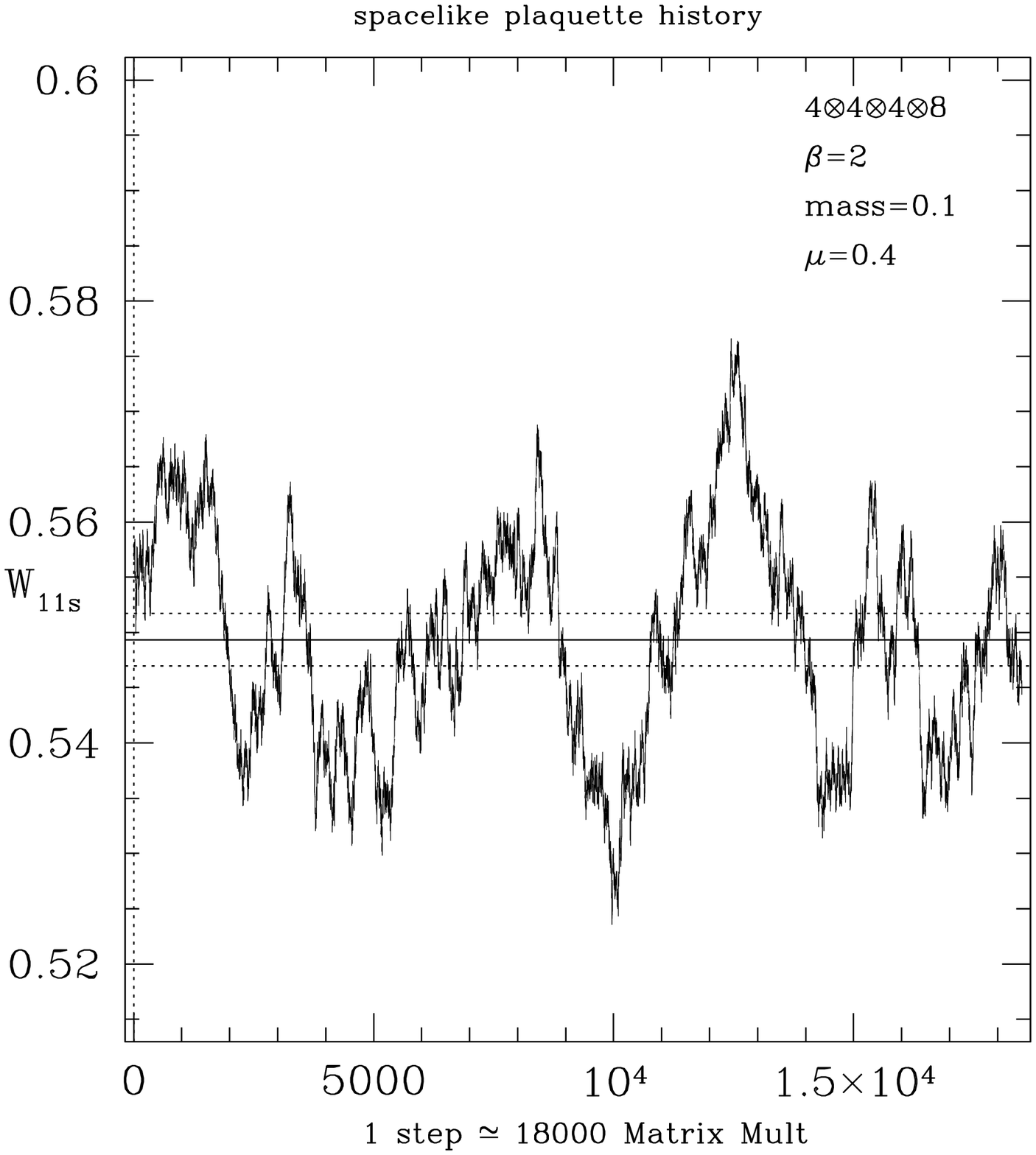
}}
\psfull
\caption{Plaquette time history for TSMB at $\mu=0.4$
\label{fig:hi_TSMB_04}}
\end{figure}

For large values of the chemical potential things get much more
difficult for both algorithms. For the representative point at
$m=0.1$, $\mu=0.4$ we cannot give a precise determination of the
integrated autocorrelations, due to the impracticability of 
accumulating sufficient statistics. However we can see two typical
histories in Figs.~\ref{fig:hi_HMC_04} and \ref{fig:hi_TSMB_04}.
The plots  look superficially similar;
however one can see from the number of matrix multiplications that 
the HMC history is much longer, consisting of approximately
15 times more matrix multiplications. This suggests an estimate of the
slowing down of HMC with respect to TSMB at that point of about
one order of magnitude. The relatively poor performance of HMC in this region 
arises from the need to reduce $d\tau$ dramatically (values as small as 0.0002
were needed at the highest $\mu$ values explored)
to maintain reasonable acceptance. These results strongly suggest that
TSMB may be the algorithm of choice in the high density region,
independent of the ergodicity considerations discussed in section
\ref{sec:algo}.

Since the integrated autocorrelation time is not available in the high
density region, and also because the determinant sign and reweighting 
factor need to be taken into account, 
in the following we will quote the jackknife error.
For purely gluonic observables such 
as the plaquette this is probably a considerable underestimate of the
true value, and possibly explains why 
the mean values we present are not really in agreement
between the two algorithms. 
Another factor which may be relevant is the 
lack of ergodicity of HMC, implying that the two algorithms may be exploring
distinct phase spaces; in subsection \ref{subs:tsmbresults} we will show that
this
is the case for some of the fermionic observables.

\subsection{Physics Results from HMC}

\begin{figure}[htb]
\psdraft
\centerline{
\setlength\epsfxsize{400pt}
\epsfbox{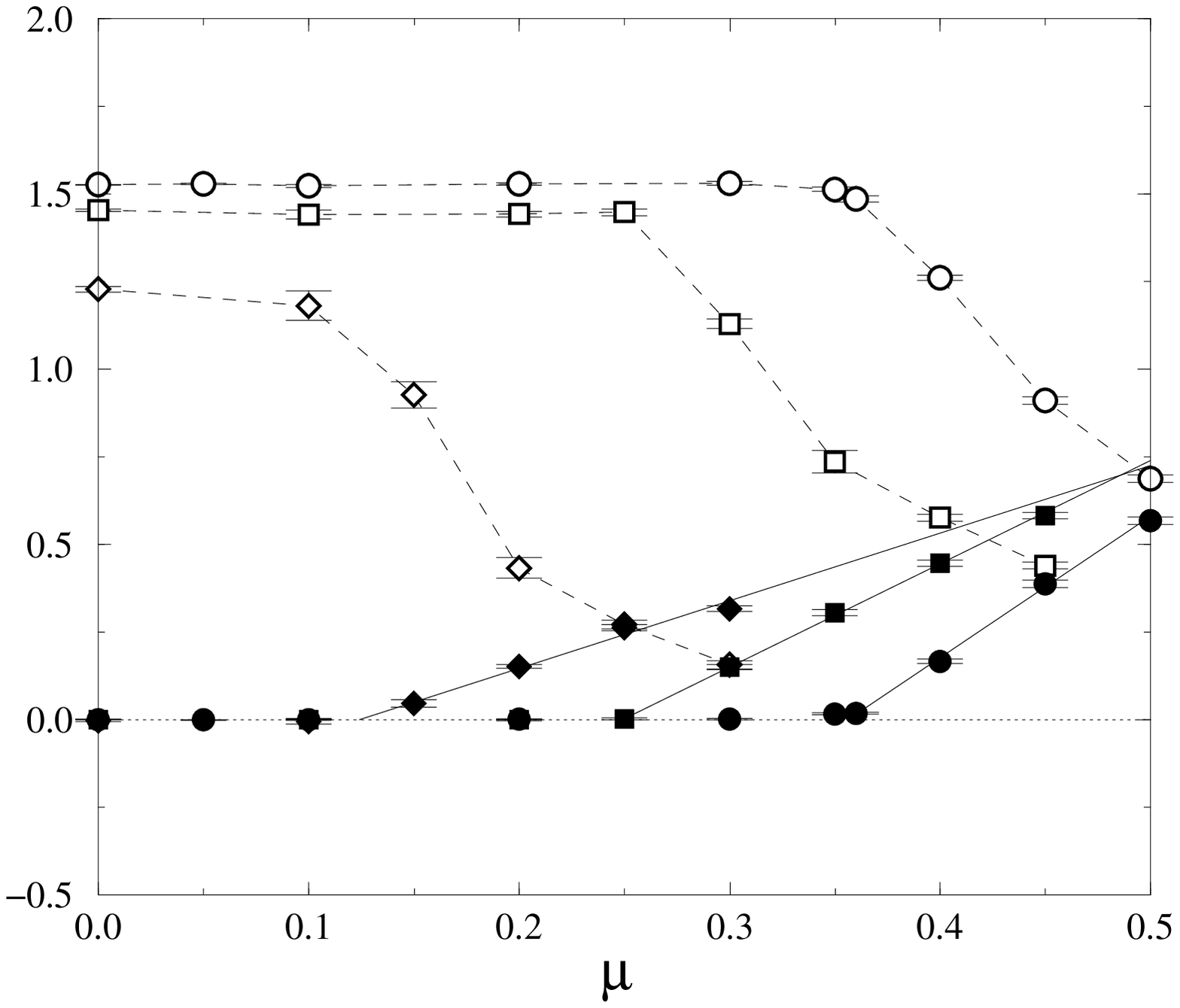
}}
\psfull
\caption{$\langle\bar\chi\chi\rangle$ (open symbols) and $n$ (filled symbols)
vs. $\mu$, for masses $m=0.1$ (circles), $m=0.05$ (squares) and $m=0.01$
(diamonds). 
\label{fig:conds}}
\end{figure}

Next we report on the results of simulations performed using the HMC
algorithm for $\mu\not=0$. We used three distinct quark masses on 
the $4^3\times8$ lattice at $\beta=2.0$, 
and explored values of $\mu$ up to and including 0.5 
for $m=0.1$, $\mu=0.45$ for $m=0.05$, and $\mu=0.3$ for $m=0.01$. 
Our results for the chiral condensate $\langle\bar\chi\chi\rangle$ 
and the baryon number density $n$ (\ref{eq:n}) are shown as a 
function of $\mu$ in Fig. 
~\ref{fig:conds}.

The results for $m=0.1,0.05$
show the condensates remaining unchanged as $\mu$ increases from 
zero up to a rather sharply-defined $\mu_o$, which we identify with the 
onset transition discussed in section \ref{subs:form}. At this point the
chiral condensate begins to fall sharply from its 
zero-density value, and the baryon density begins to rise linearly from zero.
The results for $m=0.01$, while less clear-cut, are consistent with this
picture. 
The lines through the filled points are a straight line fit to the 
non-zero values, the details of which are given in Table \ref{tab:chipt}.
The onset value $\mu_o$, corresponding to the $x$-intercept of the fit,
coincides quite well with the value of $\mu$ 
for which the edge of the eigenvalue spectrum of $M(\mu)$ crosses the imaginary axis, as illustrated in Fig.~\ref{fig:whole_spec}. 

\begin{table}[ht]
\setlength{\tabcolsep}{1.5pc}
\caption{Linear fits to $n(\mu)$ in the vicinity of the onset transition}
\label{tab:chipt}
\begin{tabular*}{\textwidth}{@{}l@{\extracolsep{\fill}}lllll}
\hline
$m$ & &$x$-Intercept & $x$-Intercept & Slope &
Slope \\
    & &($\chi$PT) & (measured) & ($\chi$PT) & (measured) \\
\hline
0.10 & &0.3664(2) & 0.356(8)  & 4.548(6)  & 3.96(6)  \\
0.05 & &0.2649(8) & 0.249(5)   & 4.141(28) & 2.94(4) \\
0.01 & &0.1271(13)& 0.117(11)  & 3.044(69) & 1.84(8)\\
\hline
\end{tabular*}
\end{table}
\begin{figure}[htb]
\psdraft
\centerline{
\setlength\epsfxsize{400pt}
\epsfbox{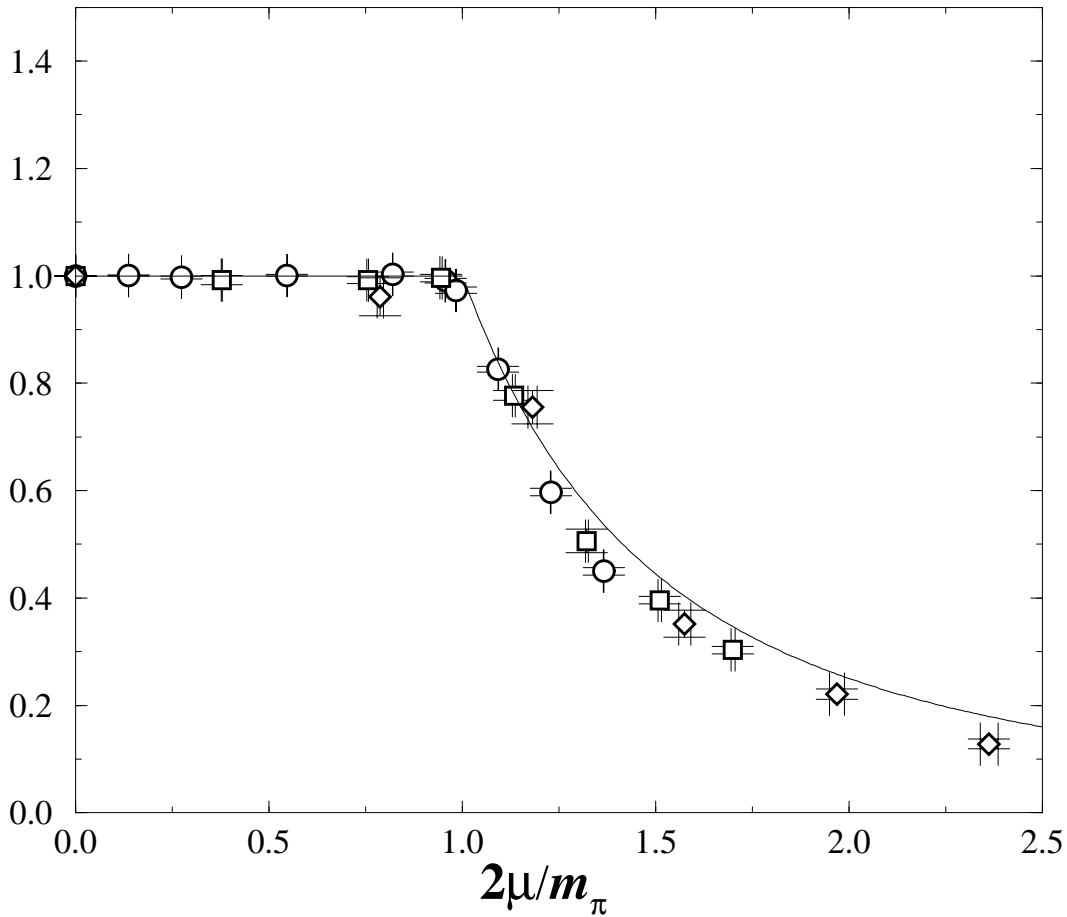
}}
\psfull
\caption{Chiral condensate vs. chemical potential
using the rescaled variables of eq. (\ref{eq:rescale}). The symbols
are the same as those of Fig.~\ref{fig:conds}
\label{fig:unipbp}}
\end{figure}
The picture is qualitatively 
very similar to the predictions displayed in figure 4 of \cite{KSTVZ}.
We can make the comparison more quantitative by replotting the data
in terms of rescaled variables $x=2\mu/m_{\pi0}$, 
$y=\langle\bar\chi\chi\rangle/\langle\bar\chi\chi\rangle_0$, and
$\tilde n=m_{\pi0}n/8m\langle\bar\chi\chi\rangle_0$, where the 0 subscript 
denotes values at zero chemical potential. The predictions from $\chi$PT 
\cite{KSTVZ} are then that all data should fall on the lines
\begin{equation}
y=\cases{1&$x<1$,\cr\cr\displaystyle{{1\over x^2}}&$x>1$;\cr}
\;\;\;\;\;\;\;\;
\tilde n=\cases{0&$x<1$,\cr\cr
\displaystyle{{x\over4}\left(1-{1\over x^4}\right)}
&$x>1$.\cr}
\label{eq:rescale}
\end{equation}
Using the data of Fig.~\ref{fig:conds} and Table \ref{tab:pionmu0}
we plot $y$ vs. $x$ in Fig.~\ref{fig:unipbp} and $\tilde n$ vs. $x$ 
in Fig.~\ref{fig:uniden}.
\begin{figure}[htb]
\psdraft
\centerline{
\setlength\epsfxsize{400pt}
\epsfbox{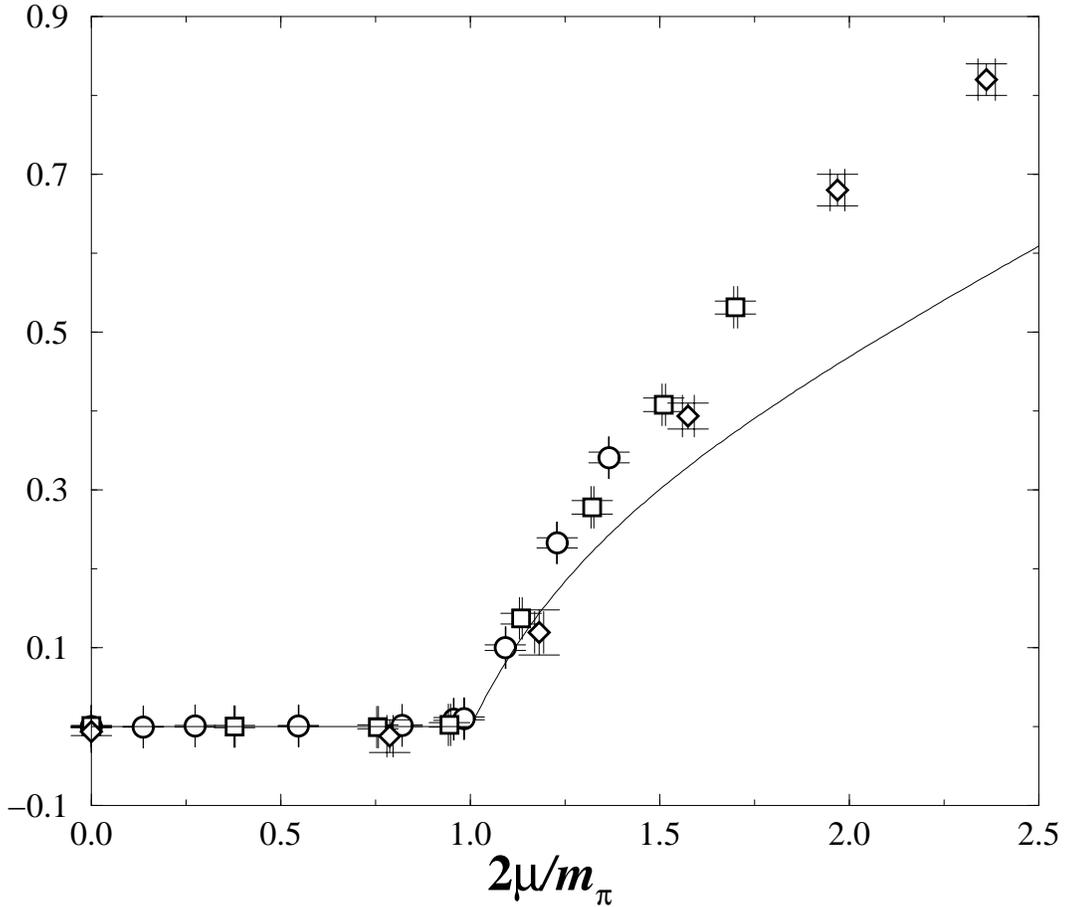
}}
\psfull
\caption{Baryon density vs. chemical potential using the rescaled variables
of eq. (\ref{eq:rescale}).
\label{fig:uniden}}
\end{figure}
\begin{figure}[htb]
\psdraft
\centerline{
\setlength\epsfxsize{400pt}
\epsfbox{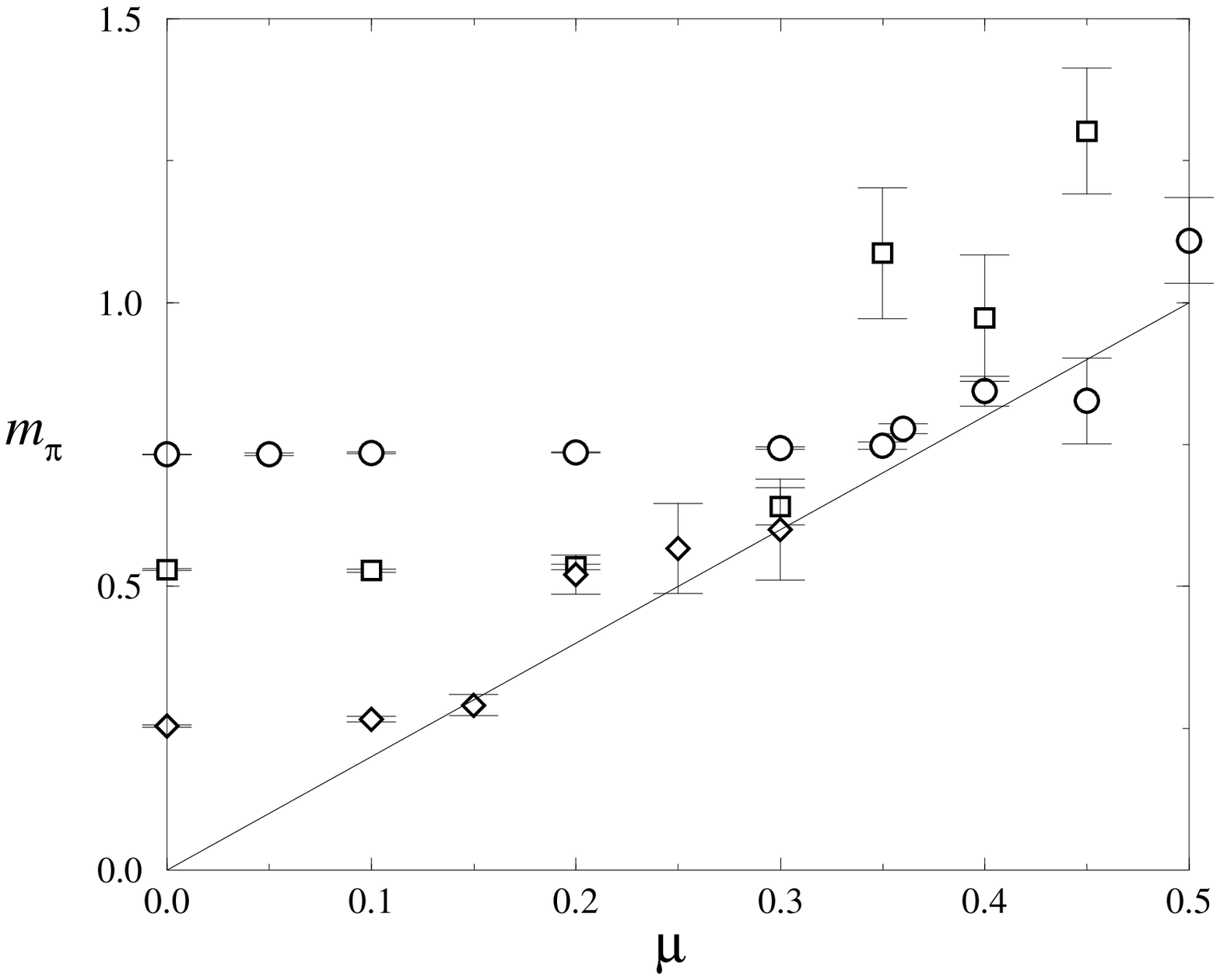
}}
\psfull
\caption{$m_\pi$
vs. $\mu$, for masses $m=0.1$ (circles), $m=0.05$ (squares) and $m=0.01$
(diamonds). Also shown is the line $m_\pi=2\mu$.
\label{fig:pion}}
\end{figure}
The data collapse very nicely onto a universal curve, corresponding
quite closely to the prediction (\ref{eq:rescale}). The systematic departures
from the theoretical curves, downwards for the condensate data and upwards
for the baryon density, may well be explicable by higher order corrections in 
$\chi$PT. In Table \ref{tab:chipt} we also list $\chi$PT
predictions for the slope and $x$-intercept of the linear fit to the 
baryon density, using a linearised approximation to (\ref{eq:rescale}):
\begin{equation}
n\simeq16{{m\langle\bar\chi\chi\rangle_0}\over m_{\pi0}^2}\left(\mu-
{m_{\pi0}\over2}\right),
\label{eq:chipt}
\end{equation}
The match between theory and measurement is excellent for the intercept
and strongly supports the identification $\mu_o\simeq m_\pi/2$; 
that for
the slope less so.
In any case, the agreement between theory and measurement is 
remarkable for data taken on such a small lattice, relatively far from 
the continuum limit, with quark masses ranging over an order of magnitude.

It seems reasonable to deduce that the high density phase for $\mu>\mu_o$
is superfluid, characterised by a non-zero diquark condensate
of the form $qq_{\bf3}$ (\ref{eq:qq3}), similar
to that observed in lattice simulations of Two Color QCD with fundamental
fermions \cite{MH}. Work to establish this by direct measurement is in progress.
Whilst the quantitative agreement between our results and the 
theoretical predictions 
of \cite{KSTVZ} is gratifying, it also contradicts the symmetry-based arguments 
of section \ref{subs:form} that there are no baryonic Goldstones for $N=1$
staggered flavor, and no gauge-invariant local diquark condensate. 
We believe that this is because the HMC simulations fail
to take into account of the determinant sign (or indeed even to change it)
ie.\ that simulations with functional weight $\vert\mbox{det}M\vert$
yield broadly similar results to those with weight $\mbox{det}^2M$; the 
premature onset at $\mu_o=m_\pi/2$
is therefore a direct manifestation of the sign and/or ergodicity problems.

Next, we investigate $m_\pi$ as a function of $\mu$. Recall that 
for $\mu\not=0$, the pion timeslice propagator is defined as
\begin{equation}
G_\pi(t)=\sum_{\vec x}M^{-1}_{0,\vec0:t,\vec x}(\mu)
M^{tr-1}_{0,\vec0;t,\vec x}(-\mu),
\label{eq:Gpi}
\end{equation}
necessitating two inversions of $M$. We implemented (\ref{eq:Gpi})
using a source site and color chosen at random and summing over sink color,
performing a simple cosh fit to the resulting $G_\pi$ over
all 8 timeslices. The fits were stable in the low-density phase, but above the
onset transition $G_\pi$ became markedly noisier and the fit less convincing,
suggesting that perhaps a different functional form is more suitable.
Our results for $m_\pi$ are shown in Fig.~\ref{fig:pion}. 
For $\mu<\mu_o$ $m_\pi$ is constant to quite high 
precision; for $\mu>\mu_o$ it begins to rise. Qualitatively
similar behaviour has been observed in the Gross--Neveu model \cite{HKimK};
however in the present case we also have the theoretical treatment
of \cite{KSTVZ}, 
which predicts that in the high density phase the state with the 
quantum numbers of the pion has mass $m_\pi=2\mu$. Whilst the large errors
in the dense phase preclude a precise comparison, the clustering of the 
points just above the line $m_\pi=2\mu$ is striking.

\begin{figure}[htb]
\psdraft
\centerline{
\setlength\epsfxsize{400pt}
\epsfbox{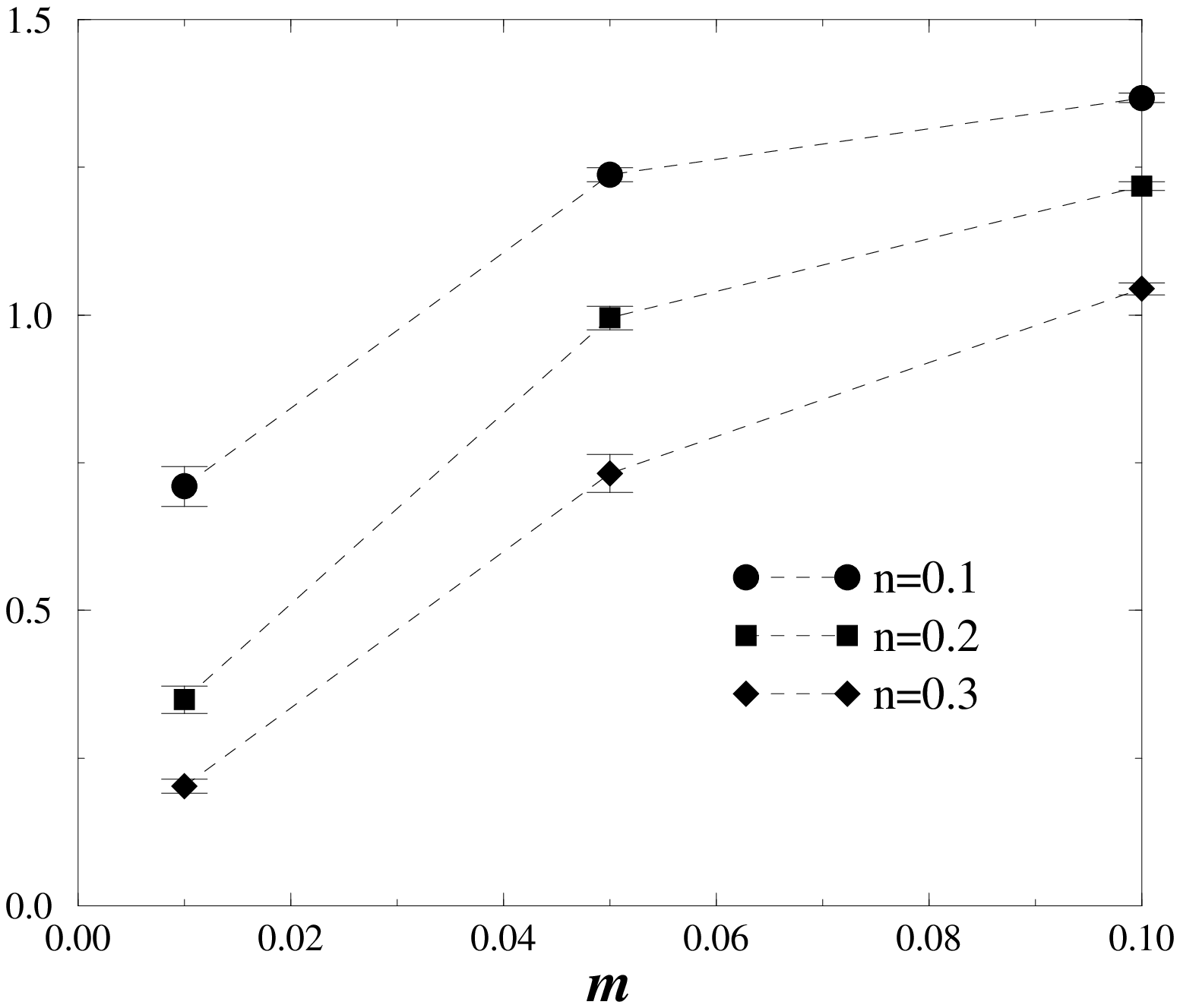
}}
\psfull
\caption{$\langle\bar\chi\chi\rangle$
vs. $m$ for various fixed baryon densities.
\label{fig:pbpvsn}}
\end{figure}

The issue of whether chiral symmetry is restored in the dense phase, 
ie.\ whether $\lim_{m\to0}\langle\bar\chi\chi(m)\rangle\not=0$, is complicated
by the sensitivity of $\mu_o$ to $m\propto m_\pi^2$. 
Eq.~(\ref{eq:rescale}) suggests
that $\langle\bar\chi\chi\rangle$ should decrease as $\mu^{-2}$ for $\mu>\mu_o$,
approaching zero 
only asymptotically as $\mu\to\infty$. This follows from the idea that 
the chiral condensate is gradually rotated into a diquark condensate as
$\mu$ increases.
Fig.~\ref{fig:unipbp} shows the first hints of this behaviour.
We can, however, take a more pragmatic (as well as more physical)
approach and plot 
$\langle\bar\chi\chi(m)\rangle$ at fixed $n$ rather than at fixed $\mu$.
This has been done for 
$n=0.1$ 0.2 and 0.3 using the linear fit to $n(\mu)$ of Table \ref{tab:chipt}
and a simple-minded linear interpolation 
of the chiral condensate data.
The result is shown in Fig.~\ref{fig:pbpvsn}.
Clearly data from more values of $m$ 
would be needed to make a definitive statement, but there
is a suggestion that chiral symmetry is not completely restored, 
particularly at the lowest density $n=0.1$. 

\begin{figure}[htb]
\psdraft
\centerline{
\setlength\epsfxsize{400pt}
\epsfbox{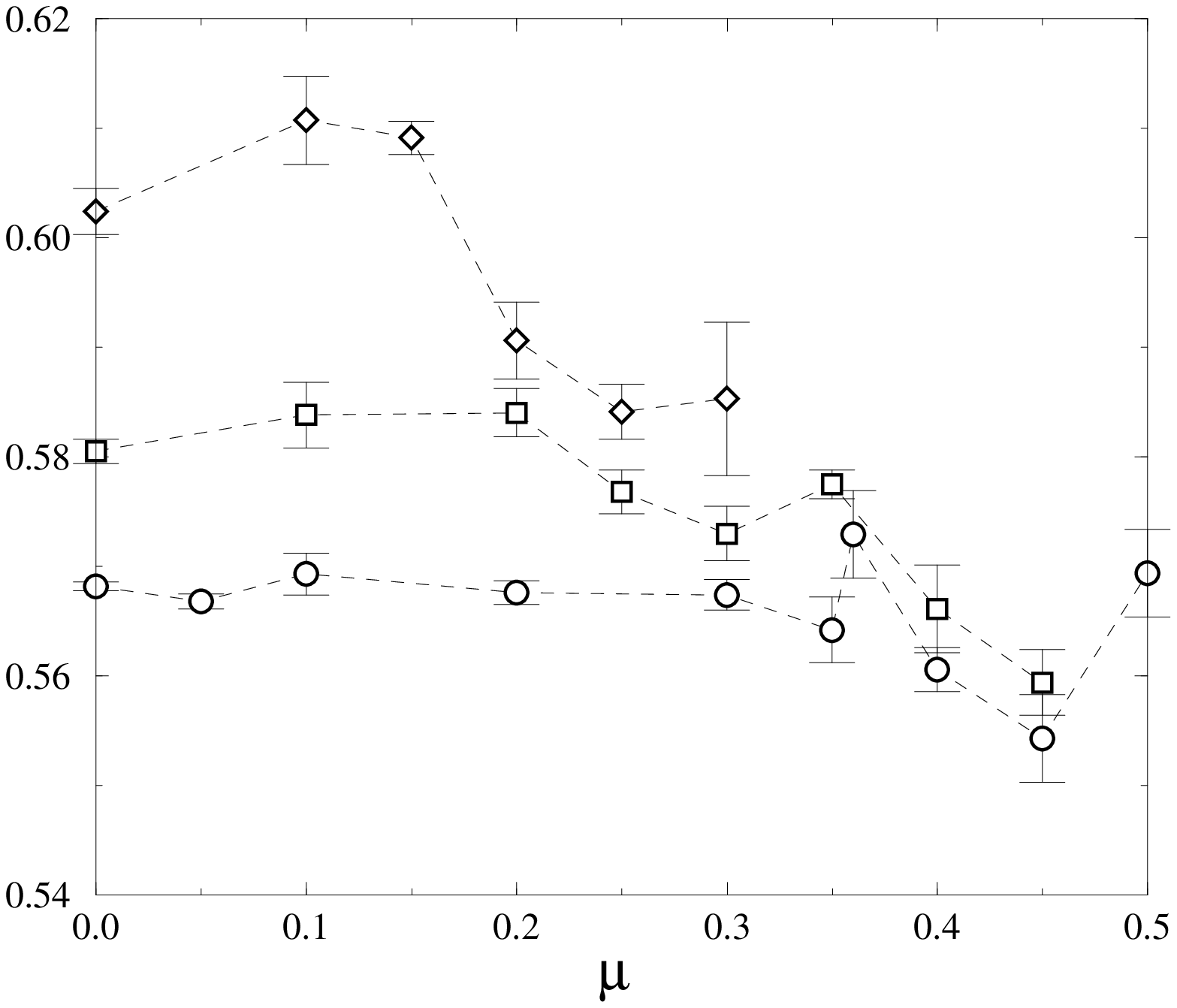
}}
\psfull
\caption{Average plaquette $\Box$
vs. $\mu$ for $m=0.1$ (circles), 0.05 (squares) and 0.01 (diamonds).
\label{fig:plaq}}
\end{figure}

Finally we turn to the effect of the chemical potential on the gauge fields.
Since this can only be communicated via fermion loops, any effect we see
can be ascribed with certainty to dynamical fermions (we cannot exclude 
the possibility that all our other 
observations could have been made for a fraction of the cost 
in the quenched approximation). Gluonic observables, however, are also
much more 
prone to auto-correlations as described in section \ref{subs:auto}, 
particularly as the quark mass is reduced. Systematic changes with $\mu$
are therefore quite difficult to observe. In this initial HMC study we 
have only measured the average plaquette; the results are shown in 
Fig.~\ref{fig:plaq}.
The data for $m=0.1$, 0.05 show the plaquette remaining roughly constant
for $\mu<\mu_o$, before beginning to decrease. The $m=0.01$ data are 
consistent with this picture within admittedly large errors. We interpret 
it as follows: for temperature $T=0$, all values of $\mu<\mu_o$ are physically
equivalent corresponding to the same physical state, namely the vacuum.
We only expect an effect on gluonic observables in the presence of matter,
ie.\ for $\mu>\mu_o$. To the extent that the results are constant for
$\mu<\mu_o$ we can be confident that our simulation has an effective 
$T\simeq0$. The decrease in the plaquette for 
$\mu>\mu_o$ may be due to the decrease in the
number of virtual quark--anti-quark pairs which may form due to the Exclusion
Principle
--- an effect known as {\sl Pauli Blocking\/}. This results
in a decrease of screening via vacuum polarisation, and hence an effective
renormalisation of the gauge coupling $\beta$ and consequent decrease of 
the plaquette. In the large-$\mu$ limit the lattice should become 
saturated with one quark of each color per site, and the plaquette assume its
quenched value; this has been verified at strong gauge coupling \cite{HM2}.

\subsection{Physics Results from TSMB}
\label{subs:tsmbresults}

Each TSMB simulation is characterised by a vector $n_i$ specifying the 
polynomial orders at each stage, as described in section \ref{subs:tsmb}.
For each configuration generated, a reweighting factor $r$ and the sign of 
$\mbox{det}M$ must be determined. On the relatively small lattices considered
here, it is possible to compute $\mbox{det}M$ directly using standard numerical
methods; on larger lattices one can use the `spectral flow method' 
\cite{DESYMUNSTER2}
as a function of $m$ and/or $\mu$.
The expectation value of an observable $O$ is then determined by the 
ratio
\begin{equation}
\langle O\rangle={{\langle O\times r\times sign\rangle}\over
                  {\langle r\times sign\rangle}}.
\label{eq:Osign}
\end{equation}
Here we present results from runs on a $4^3\times8$ lattice with $\beta=2.0$,
$m=0.1$ at three values of chemical potential: $\mu=0.0$ with polynomials
$n_i=(64,250,300)$; $\mu=0.36$ with both $n_i=
(64,300,350,500)$ and $n_i=(80,300,350,500)$; and $\mu=0.4$
with $n_i=(160,1000,1100,1200)$ (reweighting was not performed 
at $\mu=0.0$). Note that the polynomial orders required increase
with $\mu$. The point at $\mu=0$
was chosen to enable the TSMB algorithm to be tested against HMC, since 
both should yield identical results. The $\mu\not=0$ points were
chosen so as to have one value just past the HMC onset transition,
where the edge of the eigenvalue distribution just overlaps the line
$\mbox{Re}\lambda=0$ and a small percentage of negative determinant
configurations are expected, so that hopefully the sign problem is not too
severe, and one value fairly deep in the high density 
phase. 
 
Our results for the standard observables, together 
with the corresponding HMC results, are summarised in Table
\ref{tab:tsmb}. For TSMB at $\mu\not=0$ we also include 
observables determined separately in each sign sector, defined by
$\langle O\rangle_\pm=\langle O\times r\rangle_\pm/\langle r\rangle_\pm$.
The observables at $\mu=0.36$ and 0.4 result from runs on 32 separate
configurations with a few thousand update cycles each. At $\mu=0.0$ the
results quoted for $\langle\bar\chi\chi\rangle$ and $n$ come from a run on
128 configurations with $\sim$1500 update cycles. Autocorrelation studies
reveal that these may not be fully thermalised or decorrelated, which may
account for the slight discrepancies between TSMB and HMC. The plaquette
at $\mu=0.0$, by far the slowest observable to decorrelate, is taken from a
long run of $\sim$25000 update cycles on a single configuration with the same
$n_i$, 
with measured autocorrelations taken into account, but ignoring 
reweighting (reweighting, including $\mbox{sign}(\mbox{det}M)$ has not been 
observed to have a significant effect on the plaquette).
For $\mu=0.0$ and 0.4 the second
number quoted results from long runs with
40000 update cycles at $n_i=(24,90,120)$ and 
19000 at $n_i=(140,1000,1100)$ --- these runs have been used to 
obtain Figs.~\ref{fig:ac_TSMB_00} and \ref{fig:hi_TSMB_04}
respectively. Even so, at $\mu=0.4$
the long autocorrelation time implies that the errors are likely to be
underestimated, which perhaps explains the large discrepancy between 
the plaquette value including the measured autocorrelation and the other
plaquette values (note also 
that the HMC plaquette value at $\mu=0.36$ appears to lie
outside the trend of Fig.~\ref{fig:plaq}). 

\begin{table}\label{tab:tsmb}
\begin{tabular}{llrrrr} \hline
 & $\mu$ & \multicolumn{3}{c}{TSMB} & HMC \\ \cline{3-5}
 & & $\langle O\rangle$ & $\langle O\rangle_+$ & $\langle O\rangle_-$ & 
\\ \hline
$\langle\bar\chi\chi\rangle$ & 0.0 & 1.510(4) & & & 1.526(1) \\
 & 0.36 & 1.562(18) & 1.538(15) & 1.240(58) & 1.485(9) \\
 & 0.4  & 1.26(15) & 1.24(4) & 1.23(4) & 1.253(10) \\ \hline


$n$ & 0.0 & -0.0018(26) & & & -0.0002(3) \\
 & 0.36 & -0.005(14) & 0.015(12) & 0.263(52) & 0.0172(28) \\
 & 0.4  & 0.09(13) & 0.17(3) & 0.21(3) & 0.1667(90) \\ \hline

$\Box$ & 0.0 & 0.5765(58) & & & 0.5682(4) \\
 & & 0.5667(17) \\
 & 0.36 & 0.5541(12) & 0.5542(11) & 0.5551(18) & 0.5729(40)\\
 & 0.4 & 0.588(7) & 0.5791(14) & 0.5746(15) & 0.5612(30) \\ 
 & & 0.5523(46) \\ \hline
\end{tabular}
\caption{A comparison of results between TSMB and HMC
}
\end{table}
\begin{figure}[htb]
\psdraft
\centerline{
\setlength\epsfxsize{400pt}
\epsfbox{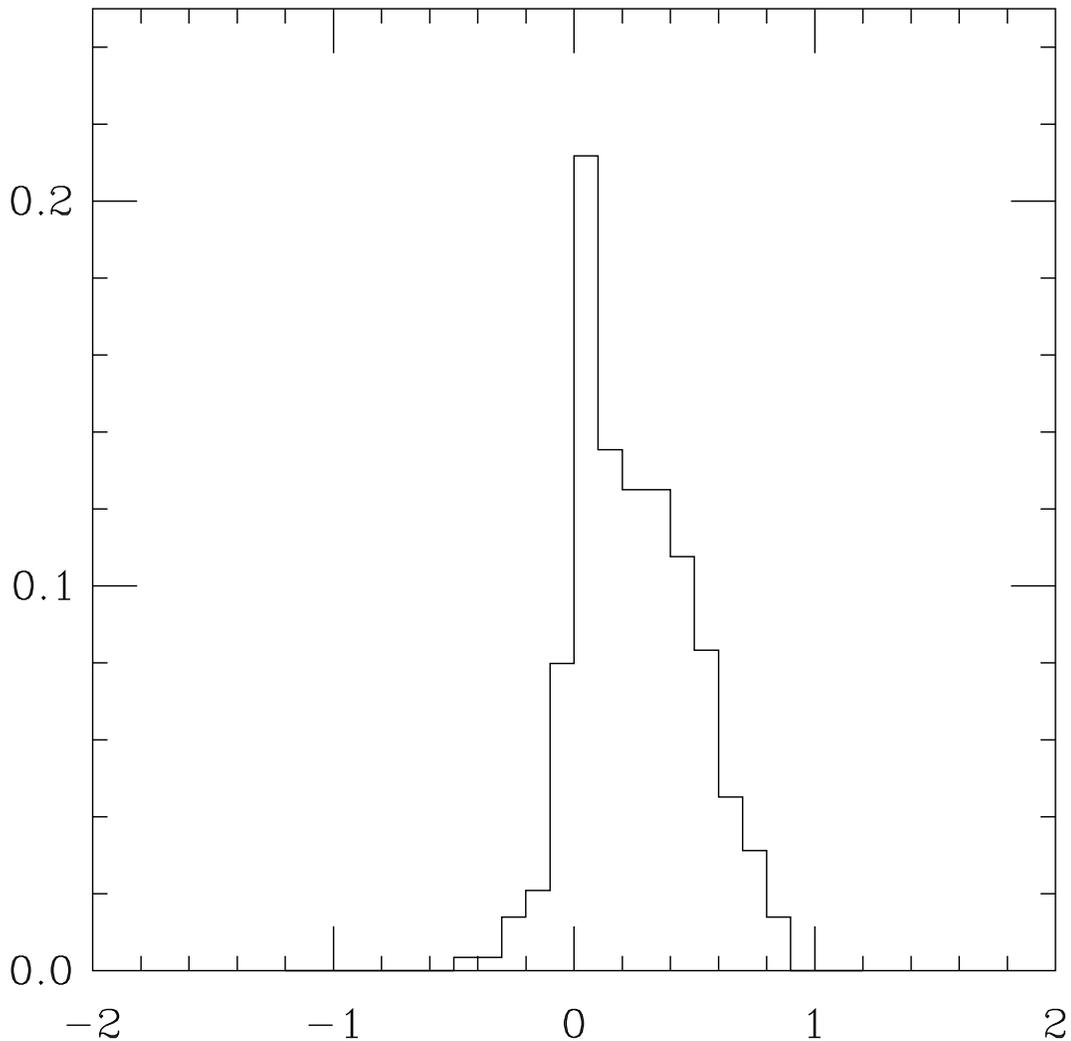
}}
\psfull
\caption{Histogram of $r\times sign$ for $\mu=0.36$ with 
$n_i=(64,300,350,500)$.
\label{fig:hist36}}
\end{figure}
\begin{figure}[htb]
\psdraft
\centerline{
\setlength\epsfxsize{400pt}
\epsfbox{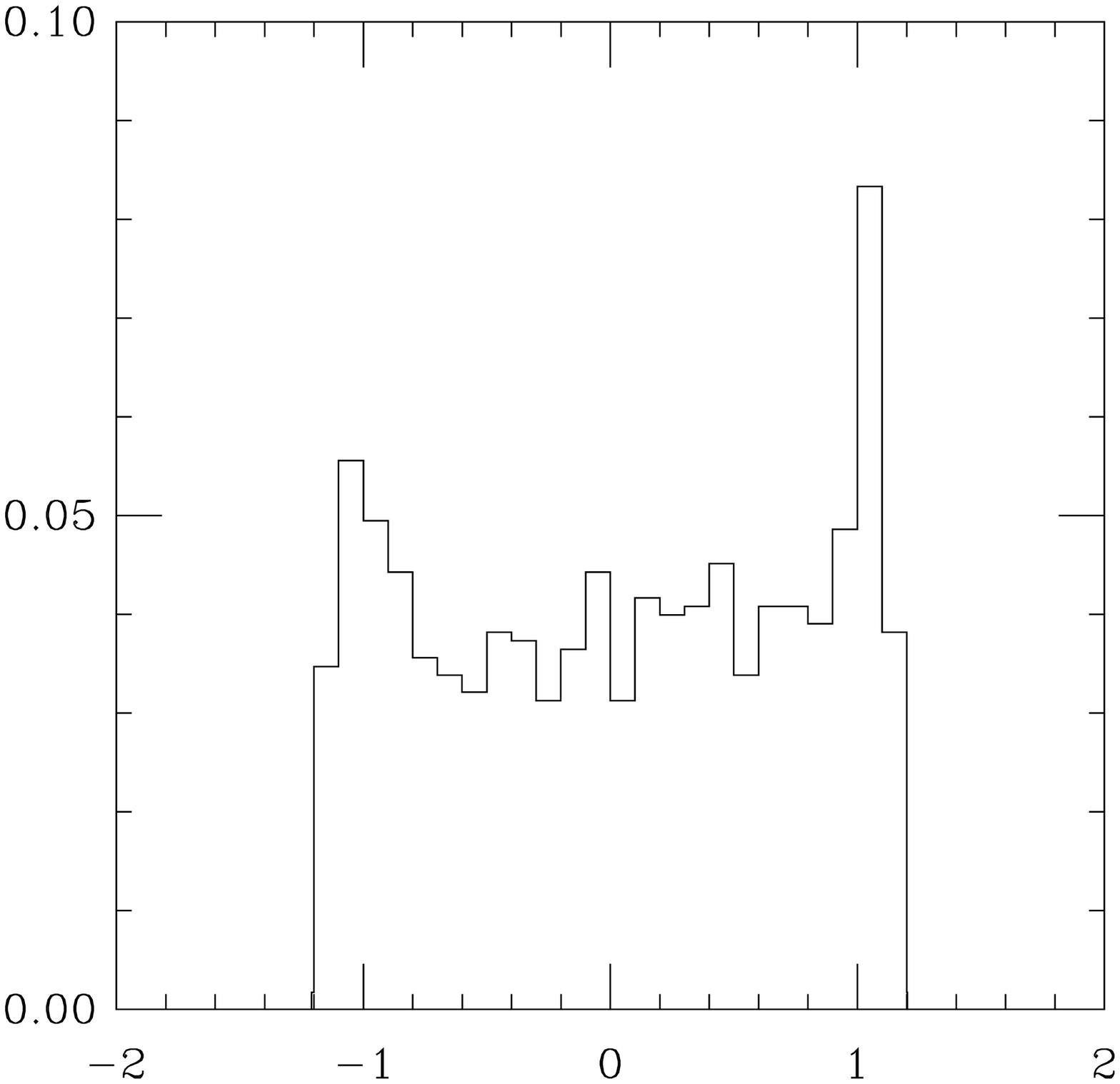
}}
\psfull
\caption{Histogram of $r\times sign$ for $\mu=0.40$ with 
$n_i=(200,1000,1100,1200)$.
\label{fig:hist40}}
\end{figure}

At $\mu=0.0$ the plaquette results in the second row 
provide reassurance that the two algorithms agree, and that we have some 
degree of control over errors. Only once $\mu>\mu_o$ (as determined by HMC)
do we anticipate the two should differ 
as a result of HMC's failure to explore the
negative determinant sector. In Figs.~\ref{fig:hist36} and \ref{fig:hist40}
we plot distributions of $r\times sign$ obtained at $\mu=0.36$, 0.4
respectively.
For $\mu=0.36$ there is a small fraction
of negative determinant configurations; for $\mu=0.4$ the distribution
is much more symmetrical between + and --, 
although the precise shape is found to be sensitive to the
choice of $n_i$, and $n_2$ in particular. 
The severity of the sign problem in numerical simulations
is usually expressed in terms of the average sign $\langle sign\rangle$; 
generically this decreases towards zero as the system volume 
increases until its relative
error becomes so large that accurate estimates of $\langle O\rangle$ are
impracticable. For TSMB the corresponding quantity is not uniquely defined;
eg.\ for the data of Fig.~\ref{fig:hist36} we could specify the fraction
of negative determinant configurations (12\%), $\langle r\times
sign\rangle/\langle r \rangle=0.910$, or $(\langle r\rangle_+-\langle r
\rangle_-)/(\langle r\rangle_++\langle r\rangle_-)=0.491$.

A signal for physical effects 
associated with the inclusion of the sign of the determinant in
the functional measure following (\ref{eq:Osign}) is that
$\langle O\rangle_+\not=\langle O\rangle_-$. The centre columns
of table \ref{tab:tsmb} reveal evidence for such an effect in the 
fermionic obsevables for $\mu=0.36$. As a result, the full average over
both sectors for $\langle\bar\chi\chi\rangle$ is significantly greater 
than the HMC result, and perhaps even consistent with
$\langle\bar\chi\chi\rangle_0$. More spectacularly, the average baryon density
$n$ is consistent with zero. These observations imply that at this value of
$\mu$ the system is still in the low density phase, and hence 
$\mu_{o\,TRUE}>\mu_{o\,HMC}$. This is
in accord with our symmetry-based
arguments that for $N=1$ flavors of adjoint staggered fermion there are no
baryonic Goldstones and hence no early onset. Note that similar effects seen
at $\mu=0.4$ are statistically far less significant, and should at this stage
be considered preliminary.

We have seen that TSMB simulations give evidence for a delayed onset,
giving us confidence that the algorithm 
correctly samples the two sign sectors and thus correctly describes a single
flavor. This is the principal result of our initial TSMB studies.

\section{Conclusions}

To our knowledge this has been the first study of Two Color QCD with adjoint
quarks, and the first TSMB study both to use staggered fermions, and 
to set $\mu\not=0$.
The highlights of this work are the following:

\begin{itemize}

\item 
We have outlined the global symmetries and anticpated patterns of 
symmetry breaking for SU(2) lattice gauge theory with staggered fermions 
and a non-zero chemical potential in both
fundamental and adjoint representations. The case of $N=1$ adjoint flavor seems
especially interesting, being the most `QCD-like'.

\item
We have studied the model using both conventional
hybrid Monte Carlo (HMC) and Two-Step Multi-Bosonic (TSMB) algorithms.
The HMC simulations slow down dramatically, both in terms of number of 
matrix inversions required, and in terms of autocorrelation, 
once the eigenvalue distribution
includes some with negative real part, which begins to occur once
baryon density $n>0$. We have confirmed the existence of 
isolated real negative eigenvalues at large $\mu$ and hence a sign problem.
The HMC algorithm is unable to change the sign of the determinant, and 
hence is not ergodic in this region.
The TSMB algorithm, by reason of the approximate way in which it treats 
small eigenvalues, is able to change the determinant sign --- our first
studies also indicate it decorrelates more effectively than 
HMC in the dense phase.

\item
Simulations using HMC over a range of $m$ and $\mu$ show good quantitative
agreement with a chiral perturbation theory treatment 
--- data spanning an order of 
magnitude in quark mass collapsing onto a simple universal curve 
whose parameters are completely determined by physical quantities 
measured at $\mu=0$. In particular a second order phase transition
(see Fig.~\ref{fig:uniden}) 
at $\mu=m_\pi/2$ separates the vacuum from a phase with $n>0$, which can 
be interpreted as a fluid of mutually repelling diquark bosons \cite{KSTVZ}.
The $\chi$PT analysis should only hold for the global 
symmetries of the model with $N\ge2$ flavors, implying that the failure to 
explore the negative determinant sector of the model results in the `wrong' 
physics.

\item
Measurements made using TSMB simulations which take the 
determinant sign into account give evidence for a delayed onset,
ie.\ $n$ remains consistent with zero
 even for $\mu>m_\pi/2$. This is in accord with the
symmetry breaking pattern anticipated for $N=1$.

\end{itemize}

In the future we plan to extend our simulations of the dense phase
using both algorithms, with the 
following goals:

\begin{itemize}

\item
We wish to examine the spectrum of the model at $\mu=0$ in greater detail,
in order to establish the masses of baryonic states such as vector diquarks,
$qg$ fermions, etc, which may be associated with further thresholds as 
they are induced into the ground state as
$\mu$ is raised. We expect the $\chi$PT treatment, which does  not include
such states, to cease to be accurate at some point.
This may be signalled either by the breakdown of universality of data with
different $m$, or by further phase transitions.

\item
We wish to probe larger values of $\mu$, up to the saturation point 
$n\simeq3$ quarks per lattice site, to see whether any new phases emerge.
Perhaps at some point a sharp Fermi surface will appear.

\item
We wish to explore gluodynamics in the dense medium \cite{Lombardo}, by studying (in order
of sophistication) Pauli Blocking, the static
quark potential, the gluon propagator, and the spatial and size distribution of 
instantons. These studies will require a 
sizeable increase in lattice volume.

\item
We wish to explore signals for diquark condensation in the dense phase,
both in superfluid and superconducting channels. This latter exotic phenomenon
will require good quantitiative control over the determinant sign fluctuations,
and hence high statistics (though probably not a large volume).

\end{itemize}

\section*{Acknowledgements}

This work is supported
by the TMR network ``Finite temperature phase transitions in particle physics''
EU contract ERBFMRX-CT97-0122. SJH thanks the Institute for Nuclear Theory
at the University of Washington for its hospitality and the Department of Energy
for partial support during the completion of this work.
Numerical work was performed using a Cray T3E
at NIC, J\"ulich and an SGI Origin2000 in Swansea.
We are grateful to Ian Barbour and Dominique Toublan
for stimulation and encouragement.

\end{document}